\documentclass[twocolumn,english,superscriptaddress,
preprintnumbers,amsmath,amssymb,floatfix,prx]{revtex4}

\usepackage[T1]{fontenc}
\usepackage[latin9]{inputenc}
\usepackage{dcolumn}
\usepackage{bm}
\usepackage{graphicx}
\usepackage{color}
\usepackage{esint}
\usepackage{babel}
\usepackage{bbold}
\usepackage{amsfonts}
\usepackage{slashed}
\usepackage{bm}
\usepackage{amsmath}
\usepackage{epsfig}
\usepackage{enumerate}
\def\be{\begin{equation}}
\def\ee{\end{equation}}
\def\bea{\begin{eqnarray}}
\def\eea{\end{eqnarray}}
\def\bmat{\begin{pmatrix}}
\def\emat{\end{pmatrix}}
\def\bs{\begin{split}}
\def\es{\end{split}}

\def\~{$\approx$}

\def\dag{\dagger}

\begin{document}

\title{Boundary Green's function approach for spinful single-channel and multichannel Majorana nanowires}

\author{M. Alvarado}
\affiliation{Departamento de F{\'i}sica Te{\'o}rica de la Materia Condensada C-V, Condensed Matter Physics Center (IFIMAC) and Instituto Nicol\'as  Cabrera,  Universidad Aut{\'o}noma de Madrid, E-28049 Madrid, Spain} 
\author{A. Iks}
\affiliation{Institut f\"ur Theoretische Physik, Heinrich-Heine-Universit\"at, D-40225  D\"usseldorf, Germany}
\author{A. Zazunov}
\affiliation{Institut f\"ur Theoretische Physik, Heinrich-Heine-Universit\"at, D-40225  D\"usseldorf, Germany}
\author{R. Egger}
\affiliation{Institut f\"ur Theoretische Physik, Heinrich-Heine-Universit\"at, D-40225  D\"usseldorf, Germany}
\author{A. Levy Yeyati}
\affiliation{Departamento de F{\'i}sica Te{\'o}rica de la Materia Condensada C-V, Condensed Matter Physics Center (IFIMAC) and Instituto Nicol\'as  Cabrera,  Universidad Aut{\'o}noma de Madrid, E-28049 Madrid, Spain} 

\date{\today}

\begin{abstract}
The boundary-Green's function (bGF) approach has been established as a powerful theoretical technique for computing 
the transport properties of tunnel-coupled hybrid nanowire devices.  
Such nanowires may exhibit topologically nontrivial superconducting phases with Majorana bound states at 
their boundaries.   We introduce a general method for computing the bGF of spinful multichannel lattice models 
for such Majorana nanowires, where the bGF is expressed in terms of the roots of a secular polynomial evaluated in complex momentum space. 
In many cases, those roots, and thus the bGF, can be accurately described 
by simple analytical expressions, while otherwise our approach allows for the numerically efficient evaluation of bGFs. 
We show that from the behavior of the roots, many physical quantities of key interest can be inferred, e.g., the value of bulk topological invariants, the energy dependence of the local density of states, or the spatial decay of subgap excitations. We apply the method to single- and two-channel nanowires of symmetry class D or DIII.
In addition, we study the spectral properties of multiterminal Josephson junctions made out of such Majorana nanowires.  
\end{abstract}
\maketitle

\section{Introduction}\label{sec1}

The interest in proximitized nanostructures where topological superconductor  phases could be engineered
is continuing to grow \cite{Alicea2012,Leijnse2012,Beenakker2013,Sarma2015,Aguado2017,Sato2017,Lutchyn2018,Zhang2019}. In particular, the case of
one-dimensional (1D) semiconducting hybrid nanowires with strong Rashba spin-orbit interaction has been intensely studied as a potential route towards the generation of Majorana bound states (MBSs)   
 \cite{Mourik2012,Albrecht2016,Deng2016,Nichele2017,Suominen2017,Gazi2017,Zhang2018,Fornieri2019,Vaiti2019}.  Such states are of high interest for  
topological quantum information processing applications \cite{Sarma2015}. 
While a phase with broken time-reversal symmetry (class D) can be expected for the cited nanowire experiments because of the presence of a magnetic Zeeman field (we 
use the abbreviation `TS' for such topological superconductors below), 
a time-reversal invariant topological superconductor (TRITOPS) phase has been predicted from related wire constructions  \cite{Wong2012,Nakosai2012,Deng2012,Zhang2013,Keselman2013,Dumitrescu2013,Chung2013,Nakosai2013,Haim2014,Gaud2014,Klinovaja2014,Schrade2015,Haim2016,Kim2016,Mellars2016,Camjayi2017,Schrade2018,Haim2019a}.
The TRITOPS phase has symmetry class DIII and is still awaiting experimental tests.  
The interest in hybrid nanowires goes well beyond the generation of topological phases. 
For instance, recent microwave spectroscopy experiments have investigated the role of spin-orbit coupling effects on the formation of Andreev
bound states \cite{Tosi2019}. 

The physics of devices made from different types of nanowires coupled by tunneling contacts
has been explored by a variety of theoretical models and techniques \cite{Alicea2012,Leijnse2012,Beenakker2013,Aguado2017}.
On one hand, minimal models restrict the Hilbert space to include only subgap bound states. This key simplification then allows 
for analytical progress (see, e.g., Refs.~\cite{Flensberg2010,Leijnse2011} for early contributions).
On the other hand, microscopic models aim for a more detailed understanding of how material properties can influence  transport observables  
(see, e.g., Refs.~\cite{DasSarma2012,Prada2012,Rainis2013,Stanescu2013,Cayao2015,Stanescu2018,Huang2018,Reeg2018}). Recent works along this line have studied the electrostatic potential profile along the nanowire 
\cite{Mikkelsen2018,Antipov2018,Escribano2019} and the effects of disorder on the phase diagram \cite{Winkler2019,Haim2019}. 
However, the solution of such microscopic models requires information about many model parameter values
and generally can be obtained only by performing a detailed numerical analysis.
In this context, theoretical approaches of \emph{intermediate} complexity are of high interest.
Such a framework allows one to describe transport properties by taking into account both subgap and 
continuum states while keeping the algebra sufficiently simple so as to permit analytical progress. 
The scattering matrix formalism is a widely known representative for this type of approach (see, e.g., Refs.~\cite{Akhmerov2011,Pientka2012,Houzet2013,Ioselevich2016,Setiawan2017,Reeg2017,Nakhmedov2019}).
The present paper will employ the complementary 
 boundary Green's-function (bGF) method \cite{Zazunov2016,Peng2017,Zazunov2017,Komnik2017,Zazunov2018, Istas2019}, which is 
particularly useful for analyzing nonequilibrium transport properties in different types of hybrid nanojunctions.  The bGF approach also allows one to  
examine other electronic properties such as the tunneling density of states (DoSs) or the bulk-boundary correspondence expected for topological phases \cite{Essin2011,Zhang2010,Tewari2012}.
Furthermore, electron-phonon and/or electron-electron interaction effects can in principle also be taken into account.

In the present paper, we extend and generalize the bGF approach for 1D or quasi-1D proximitized nanowires, which has been introduced in Refs.~\cite{Zazunov2016,Peng2017,Zazunov2017,Komnik2017,Zazunov2018}, along several directions. 
First, we demonstrate that a bGF construction in terms of the roots of a secular equation extended to complex momenta (as discussed in Ref.~\cite{Zazunov2016} for the Kitaev chain model) can be generalized to arbitrary spinful multichannel (i.e., quasi-1D) nanowires with topologically nontrivial superconducting phases.  
In particular, by studying the evolution of the roots in the complex momentum plane under the variation of model parameters, one can readily detect topological transitions,
determine bulk topological invariants, or compute the local density of states as a function of energy for translationally invariant cases.
In addition, the same roots determine the bGF and thereby give access to the transport properties of devices made from tunnel-coupled (semi-infinite or finite-length) nanowires.
In particular, their knowledge also gives access to the spatial decay profile of Majorana states.   

Below we  investigate the roots and the corresponding bGFs for two widely used spinful single-channel
nanowire models harboring topologically nontrivial phases.  
First, we study TS wires with broken time-reversal invariance using the model by Lutchyn \emph{et al.}~\cite{Lutchyn2010} and by Oreg \emph{et al.}~\cite{Oreg2010}.  
Second, we consider TRITOPS wires using the model of Zhang \emph{et al.}~\cite{Zhang2013}. 
Quasi-1D multichannel models in class D or class DIII are then constructed by coupling several wires of the respective symmetry class by tunnel couplings.
We show that also such multichannel models can be efficiently tackled by our bGF method.  As application, we will discuss the Josephson current-phase relation both
for a multiterminal junction composed of three tunnel-coupled TS wires and for a TRITOPS-TS Josephson junction.

The remainder of this article is organized as follows. In Sec.~\ref{sec2}, we describe a general formalism for analyzing 
1D or quasi-1D lattice models of proximitized nanowires, where we only assume that the hopping amplitudes in the corresponding tight-binding model are of finite range. We show that the real-space bulk Green's function (GF) adopts a compact expression in terms of the roots of the secular polynomial of the bulk Hamiltonian  extended into complex momentum space. We also show how the boundary GF can be obtained from the bulk GF by solving a Dyson equation, and we discuss general properties of the corresponding roots. 
In Sec.~\ref{sec3}, we consider a discretized version of the single-channel class D model of Refs.~\cite{Lutchyn2010,Oreg2010}. 
We introduce a simple ansatz for the respective roots in the trivial and in the topological phase. This ansatz allows us to obtain analytical insights about the bulk spectral density and the spatial variation of MBSs. In Sec.~\ref{sec4}, we extend the analysis to a two-channel model describing two coupled class D wires, where we can study spin-orbit interaction effects in multichannel nanowires \cite{Park2017}. 
The phase diagram and the spectral density of this model  show a richer behavior than in the single-channel case. In Sec.~\ref{sec5}, we apply our methods to  single- and multichannel models for TRITOPS wires.
Finally, in Sec.~\ref{sec6}, we study the Josephson effect and the formation of Andreev bound states in  phase-biased multiterminal TS junctions and for TRITOPS-TS junctions. We finally offer some conclusions in Sec.~\ref{sec7}.  Technical details have been delegated to two appendices. 
We often use units with $\hbar=1$ and focus on the zero-temperature limit throughout.

\section{Boundary Green's function}\label{sec2}

A central aim of the present work is to construct the bGF for different hybrid nanowire models which are described by a 
bulk Hamiltonian of the form
\begin{equation}\label{BdG}
H_{\rm bulk} = \frac{1}{2} \sum_k \hat{\Psi}^{\dagger}_k \hat{\mathcal{H}}(k)\hat{\Psi}^{}_k ,
\end{equation}
corresponding to an infinitely long and translationally invariant  (quasi-)1D chain with lattice spacing $a$.
Here, $\hat{\mathcal{H}}(k)$ is an $N\times N$ Bogoliubov-de Gennes (BdG) Hamiltonian in reciprocal space, and the $\hat{\Psi}_k$ are fermionic Nambu spinor fields. 
Specific examples for these spinor fields will be given in the subsequent sections.
The number $N$ may include the Nambu index, the spin degree of freedom, and channel indices for multichannel models.
Using $\hat{\mathcal{H}}(k+2\pi/a)=\hat{\mathcal{H}}(k)$, the BdG Hamiltonian 
can be expanded in a Fourier series, $\hat{\mathcal{H}}(k)=\sum_n \hat{\mathcal{V}}_n e^{inka}$, where Hermiticity implies 
$\hat{\mathcal{V}}_{-n}=\hat{\mathcal{V}}^{\dagger}_n$. For simplicity, we here consider only models with nearest-neighbor hopping, 
$\hat{\mathcal{V}}_{n}=0$ for $|n|>1$, but the generalization to arbitrary finite-range hopping amplitudes is straightforward. 

The retarded bulk GF of the infinite chain is defined as
\begin{equation}
\hat{G}^R(k,\omega) = \left[\omega+i0^+ - \hat{\mathcal{H}}(k)\right]^{-1},
\end{equation}
where the $N\times N$ matrix structure is indicated by the hat notation.
In real space representation, the GF has the components ($j$ and $j'$ are lattice site indices)
\begin{equation}   \label{GR11}
\hat{G}^R_{jj'}(\omega) = \frac{a}{2\pi} \int_{-\pi/a}^{\pi/a} dk \, e^{i(j-j')ka}\, \hat{G}^R(k,\omega) .
\end{equation}
By the identification $z = e^{ika}$, this integral is converted into a complex contour integral, 
\begin{equation}   \label{G1}
\hat{G}^R_{jj'}(\omega) = \frac{1}{2\pi i} \oint_{|z|=1} \frac{dz}{z} z^{j-j'} \hat{G}^R(z,\omega)  .
\end{equation}
Introducing the roots $z_n(\omega)$ of the secular polynomial in the complex-$z$ plane,
\begin{equation}\label{polysec}
P(z,\omega) = \mbox{det}\left[\omega - \hat{\mathcal{H}}(z)\right]= \frac{1}{z^N}\prod_{n=1}^{2N} \left[z-z_n(\omega)\right],
\end{equation} 
the contour integral \eqref{G1} can be written as a sum over the residues of all roots inside the unit circle: 
\begin{equation}\label{cofactor}
\hat{G}^R_{jj'}(\omega)  = \sum_{|z_n|<1} \frac{z_n^{j-j'}\hat{A}(z_n,\omega) }{\prod_{m\ne n}
\left(z_n - z_m\right)},
\end{equation}
where $\hat{A}(z,\omega)$ is the cofactor matrix of $\left[\omega - \mathcal{H}(z)\right]z$.
For notational simplicity, we omit the superscript `$R$' in retarded GFs from now on.

Given the real-space components of the bulk GF in Eq.~\eqref{cofactor}, we next employ Ref.~\cite{Zazunov2016} (see also Ref.~\cite{Liliana2009}) to derive 
the bGF characterizing a \emph{semi-infinite} nanowire. To that effect, we add
an impurity potential $\epsilon$ localized at lattice site $j=0$.  Taking the limit $\epsilon\to \infty$,
the infinite chain is cut into disconnected semi-infinite chains with $j<-1$ (left side, $L$) and $j>1$ (right side, $R$).
Using the Dyson equation, the local GF components of the cut nanowire follow as \cite{Zazunov2016}
\begin{equation} \label{Dyson-n}
\hat{\mathcal{G}}_{j j}(\omega) = \hat{G}_{j j}(\omega) - \hat{G}_{j 0}(\omega) \left[\hat{G}_{00}(\omega)\right]^{-1} \hat{G}_{0 j}(\omega) .
\end{equation} 
The bGF for the left and right semi-infinite chain, respectively, are with Eq.~\eqref{Dyson-n} given by
\begin{equation}
\hat{\mathcal{G}}_L(\omega) = \hat{\cal G}_{-1,-1}(\omega) ,\quad 
\hat{\mathcal{G}}_R (\omega)=\hat{\cal G}_{11}(\omega) .
\end{equation}
We note that by proceeding along the lines of Refs.~\cite{Peng2017,Fulga2012}, one can also compute reflection matrices from the corresponding bGF, 
\begin{equation}\label{reflection_matrix}
\hat{r}_{L/R} = \lim_{\omega \to 0} \frac{\mathbb{1}-i\hat{\mathcal{V}}^\dag_{\pm1}\hat{\mathcal{G}}_{L/R}(\omega)
\hat{\mathcal{V}}_{\pm1}}{\mathbb{1}+i\hat{\mathcal{V}}^\dag_{\pm 1}\hat{\mathcal{G}}_{L/R}(\omega)\hat{\mathcal{V}}_{\pm 1}} .
\end{equation}
This relation allows one to express topological invariants of the bulk Hamiltonian \cite{Zhang2010,Tewari2012} in terms of bGFs.

The roots $z_n(\omega)$ play an important role in what follows. In particular, their 
knowledge allows us to construct both the bulk and the boundary GFs.
In simple cases, this can be done analytically, and otherwise this route offers an efficient numerical scheme.
The roots can also provide detailed information about the decay of subgap states localized at the boundaries of semi-infinite wires, and they
allow one to compute topological invariants of the bulk system.
Let us therefore summarize some general properties of these roots:
\begin{itemize}
\item[(i)] Hermiticity of the BdG Hamiltonian implies that every root $z_n(\omega)$ is accompanied by a root $1/z_n^*(\omega)$, where 
`$\ast$' denotes complex conjugation.
\item[(ii)] Electron-hole symmetry of the BdG Hamiltonian implies that $z_n(\omega)=z^*_n(-\omega)$. In the presence of an additional symmetry $\hat{\mathcal{H}}(k)=\hat{U}\hat{\mathcal{H}}(-k)\hat{U}^\dagger$ with a unitary matrix $\hat{U}$, for every root $z_n(\omega)$, also $z^*_n(\omega)$ must be a root.
\item[(iii)] As a consequence of (i) and (ii),  $\prod_{n=1}^{2N} z_n(\omega)=1$.
\item[(iv)] Topological phase transitions can occur once a pair of zero-energy roots hits the unit circle, $|z_n(0)|=1$, 
which corresponds to the closing and reopening of a gap in the bulk spectrum. 
\item[(v)] Equations~(\ref{cofactor}) and (\ref{Dyson-n}) imply that subgap bound states (with energy $E$) localized near the boundary of a semi-infinite wire  decay into the bulk 
in a manner controlled by $\max(|z_n(E)|<1)$.
\end{itemize}
We illustrate the usefulness of these properties in the following sections for different models of proximitized (quasi-)1D nanowires.
 
\section{Spinful single-channel hybrid nanowires}
\label{sec3}

As first example, we consider the spinful single-channel model of Refs.~\cite{Lutchyn2010,Oreg2010} for a proximitized semiconductor nanowire.
This model has been extensively studied as prototype for 1D wires harboring a TS phase with broken time-reversal invariance.
We use the Nambu bispinor $\hat{\Psi}_k^T=\left(c_{k\uparrow}, c_{k\downarrow},c^{\dagger}_{-k\downarrow},-c^{\dagger}_{-k\uparrow}\right)$,
i.e., $N=4$ in Eq.~\eqref{BdG}.
Here, $c_{k\sigma}$ is a fermionic annihilation operator for momentum $k$ and spin $\sigma=\uparrow,\downarrow$, and
the bulk BdG Hamiltonian in Eq.~\eqref{BdG} takes the form
\begin{equation}\label{Lutchyn}
\mathcal{H}(k) = \epsilon_k \sigma_0\tau_z + V_x\sigma_x \tau_0+\alpha_k\sigma_z\tau_z + \Delta\sigma_0 \tau_x,
\end{equation}
where $\sigma_{x,y,z}$ and $\tau_{x,y,z}$ are Pauli matrices in spin and Nambu (electron-hole) space, respectively, with the identity matrices 
 $\sigma_0$ and $\tau_0$.
Regularizing the continuum model of Refs.~\cite{Lutchyn2010,Oreg2010} by imposing a finite lattice spacing $a$,
the kinetic energy $\epsilon_k = 2t[1-\cos(ka)]-\mu$ includes the chemical potential $\mu$ and the nearest-neighbor hopping amplitude $t$.
Furthermore, $V_x$ encapsulates a magnetic Zeeman field oriented along the wire axis,
$\alpha_k =\alpha \sin(ka)$ describes the spin-orbit interaction, and $\Delta$ refers to the proximity-induced on-site pairing amplitude.
The bulk dispersion relation, $E=E_{k,\pm}\ge 0$, then follows from \cite{Lutchyn2010,Oreg2010}
\begin{equation}\label{bulkbands}
E_{k,\pm}^2 =\Delta^2 + \alpha_k^2 + V_x^2 + \epsilon_k^2 \pm 2 \sqrt{\Delta^2V_x^2 + (\alpha_k^2 + V_x^2)\epsilon_k^2}.
\end{equation}
This model exhibits a topological transition at $V_x = V_c= \sqrt{\Delta^2 + \mu^2}$, where the TS phase is realized for $V_x>V_c$.

Although it is not essential for the subsequent discussion, the parameters $t$ and $\alpha$
 can be assigned values  appropriate for InAs nanowires \cite{Zazunov2016}. To that end, we put $t = \hbar^2/(2m^* a^2)$, where $m^*$ is the effective mass, and $\alpha = \hbar u/a$, where $u$ is the spin-orbit parameter \cite{Oreg2010}.
This parameter depends on material properties and can 
be tuned by an external electric field.  Putting $a=10$~nm and using typical InAs material parameters, we estimate
$t \approx 10$~meV and $\alpha \approx 4$~meV \cite{Zazunov2016}. On the other hand, a proximity gap of order $\Delta \approx$~0.2 meV represents the case of 
a nanowire in good contact with a superconducting Al layer. (We will use this value in the figures below unless noted otherwise.)  The only remaining free variables are then given by $V_x$ and $\mu$.

Using Eq.~\eqref{polysec} and $z=e^{ika}$, the roots $z_n(\omega)$ for this model satisfy the condition
\begin{eqnarray}\label{8ord}
&&  2\Delta^2\left[\tilde\alpha^2(z) + \epsilon^2(z) - V_x^2 - \omega^2\right] \nonumber \\  
&&\, +  
2\tilde\alpha^2(z)\left[ V_x^2 - \omega^2 - \epsilon^2(z)\right] 
+ \tilde\alpha^4(z) + V_x^4 +\Delta^4 \nonumber \\ 
&&\,+ \left[\omega^2 - \epsilon^2(z)\right]^2 - 2V_x^2\left[\omega^2 +\epsilon^2(z)\right]  = 0,
\end{eqnarray}
with the  functions
\begin{equation}
    \tilde\alpha(z) = -i\alpha(z-z^{-1})/2, \quad \epsilon(z) = -t(z+z^{-1}-2)-\mu.
\end{equation}
 Equation~\eqref{8ord} can be written as
\begin{equation}
\sum_{n=1}^4 C_n(\omega) \left(z^n +\frac{1}{z^n}\right)  + C_0(\omega)= 0,
\label{polynomial}
\end{equation} 
where the real coefficients $C_n(\omega)$ are given in Appendix~\ref{appA}. Clearly, Eq.~\eqref{polynomial} is 
consistent with the general properties (i) and (ii) listed in Sec.~\ref{sec2}. Alternatively, Eq.~(\ref{polynomial}) can
be expressed as an eighth-order polynomial equation: 
\begin{equation}\label{8ord2}
    \sum_{m=0}^8 a_m(\omega) z^m =0,
\end{equation}
where the coefficients $a_m$ are trivially related to the $C_n$ and we can impose
the normalization conditions $a_0=a_8=1$.

\begin{figure}[t]
\includegraphics[width=\columnwidth]{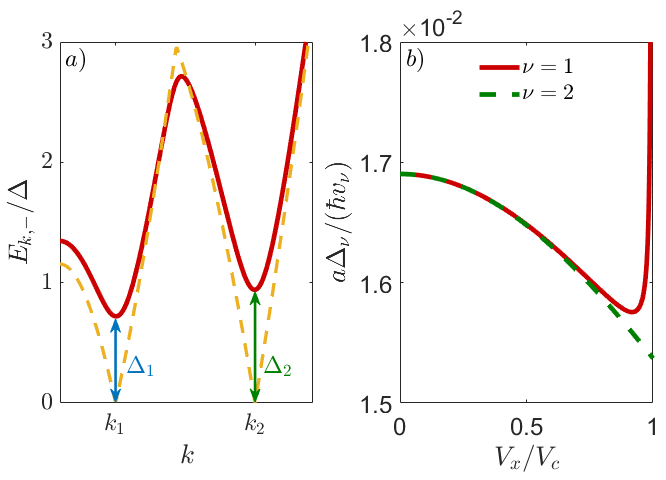}
\caption{
Bulk dispersion relation of the spinful single-channel Majorana wire model \cite{Lutchyn2010,Oreg2010}.
(a) $E_{k,-}$ vs $k$ [see Eq.~\eqref{bulkbands}] for  
 the topologically trivial regime $V_x<V_c$ (solid red curve), 
indicating the two pairing gaps $\Delta_1$ and $\Delta_2$ at $k=k_1$ and $k=k_2$, respectively [see Eq.~\eqref{k12}].
We use $\mu=5$~meV, $\Delta=2$~meV, and $V_x=0.5V_c$. All other parameters are specified in the main text. 
The dashed yellow curve is for $\Delta=0$. 
(b) Evolution of the two gaps (normalized to the velocities $v_{1,2}$) vs Zeeman parameter $V_x$ for $\Delta=0.2$~meV and $\mu=2$~meV.
We note that $\Delta_1$ and $v_1$ simultaneously vanish as $V_x\to V_c$.}
\label{fig1}
\end{figure}

The resulting roots $z_n$ can be grouped into two different classes associated with the two pairing gaps $\Delta_1$ and $\Delta_2$ in the bulk spectrum \cite{Lutchyn2010,Oreg2010} [see Fig.~\ref{fig1}(a)].
In the limit $\Delta\to 0$, these gaps $\Delta_1$ and $\Delta_2$ will also vanish. For $\Delta=0$,
we find from Eq.~\eqref{8ord} that the zero-frequency roots $z_n(\omega=0)$ simplify to $e^{\pm i k_1 a}$ and $e^{\pm i k_2 a}$, with  
\begin{eqnarray}\label{k12}
 k_{1,2} &\simeq& \cos^{-1}\Biggl( \frac{2t(2t-\mu)}{\alpha^2+4t^2}   \nonumber \\ &\pm&  
 \frac{\sqrt{V_x^2(\alpha^2 + 4t^2) + \alpha^4 + 4t\mu\alpha^2 - \alpha^2\mu^2}}{\alpha^2+4t^2} \Biggr).\quad
\end{eqnarray}
At these momenta, the dispersion relation becomes gapless for $\Delta=0$ [see Fig.~\ref{fig1}(a)].
We observe from Eq.~\eqref{k12} that $k_1$ (corresponding to the $+$ sign) becomes purely imaginary for $V_x >V_c$.  
We will then first discuss the topologically trivial regime $V_x<V_c$. 
 
Figure~\ref{fig1}(a) shows that the low-energy physics will be dominated by the regions with $|k|\approx k_1$ and $|k|\approx k_2$. 
The pairing gaps $\Delta_{1,2}=|E_{k_{1,2},-}|$ then follow by substituting $k_{1,2}$ into the bulk dispersion relation \eqref{bulkbands}.  In particular, we find that
$\Delta_1$ closes and reopens when ramping $V_x$ through the topological transition at $V_x=V_c$. 
  An approximate expression for the roots is obtained by linearizing the $\Delta=0$ dispersion relation in Eq.~\eqref{bulkbands}
for electrons and holes near $k=k_1$ and $k=k_2$.   Defining the respective velocities as $v_{\nu=1,2}=|\partial_k E_{k=k_\nu,-}|_{\Delta=0}$,  
the effective low-energy Nambu Hamiltonian valid near the respective momentum $k_\nu$ can be written as
\begin{equation}
{\cal H}_{{\rm eff},\nu=1,2} (k)\simeq \left( \begin{array}{cc}  v_{\nu}(k-k_{\nu}) & \Delta_{\nu} \\
\Delta_{\nu} & - v_{\nu}(k - k_{\nu}) \end{array}\right),
\end{equation}
and similarly for $k\approx -k_\nu$.
Using $ika=\ln z$, the condition $\det[\omega - {\cal H}_{{\rm eff},\nu}(z)]=0$ can readily be solved. In effect, the roots are given by
\begin{equation}\label{ansatz-first}
z_{\nu}(\omega)\simeq \left(1 \pm \frac{a}{v_{\nu}} \sqrt{\Delta_{\nu}^2-\omega^2}\right)e^{ ik_{\nu}a} ,
\end{equation} 
plus the complex conjugate values.
Inspired by Eq.~\eqref{ansatz-first}, we propose the following ansatz for the roots $z_n(\omega)$ located \emph{inside} the unit circle:
\begin{equation}\label{ansatz}
z_{\nu}(\omega) = \left(1-\tau_\nu\sqrt{\Delta_\nu^2-\omega^2}\right)e^{ i\delta_\nu},
\end{equation} 
where $\tau_{1,2}$ and $\delta_{1,2}$ are phenomenological coefficients.  In addition, the complex conjugate root $z^\ast_\nu(\omega)$ is a solution.
This ansatz is expected to work well in the topologically trivial regime $V_x<V_c$.
For small $\Delta$ and $|\omega|$, Eq.~\eqref{ansatz-first} implies the limiting behavior 
$\tau_{\nu}= a/ v_{\nu}$ and $\delta_\nu= k_\nu a$.
In addition, we also impose the condition
\begin{equation}\label{conditionadd}
    \tau_1 \Delta_1 = \tau_2 \Delta_2 = \eta \ll 1,
\end{equation}
where $\eta$ is a small parameter.
In the small-$\Delta$ case with $\tau_\nu\approx a/v_\nu$, Eq.~\eqref{conditionadd} implies that 
the effective pairing gap $\Delta_{\nu}$ is inversely proportional to the corresponding density of states $\propto 1/v_\nu$. 
Figure \ref{fig1}(b) shows that this condition
is accurately fulfilled as long as $V_x$ stays well below $V_c$. However, Eq.~\eqref{conditionadd} becomes
 less precise for $V_x\to V_c$.  In Appendix~\ref{appA}, we provide more refined 
analytical expressions that determine the parameters $\eta$ and $\delta_{\nu}$ in our ansatz for the roots [see Eqs.~\eqref{ansatz} and \eqref{conditionadd}].

\begin{figure}[t]
\includegraphics[width=\columnwidth]{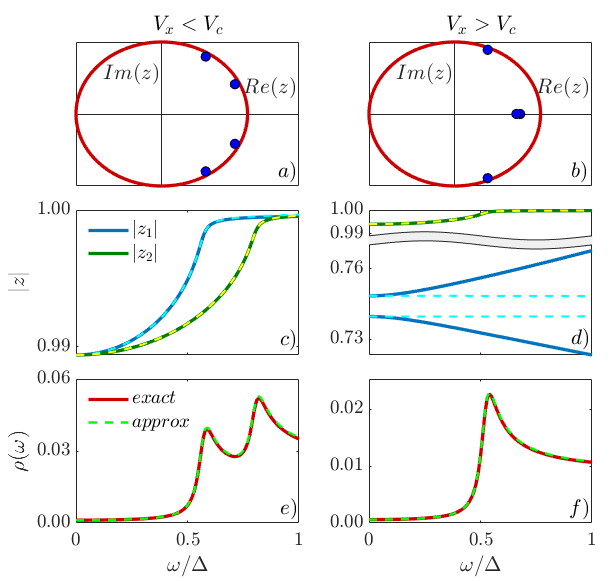}
\caption{Behavior of the roots $z_n(\omega)$ for the spinful single-channel Majorana wire model \cite{Lutchyn2010,Oreg2010}. 
We use $V_x = 0.5V_c$ and $V_x = 1.2 V_c$ as representatives for topologically trivial and nontrivial cases, respectively, 
with $\mu=5$~meV and other parameters as specified in the main text.
Upper panels: Roots $z_n(\omega=0)$ (black dots) inside the unit circle (red) for (a) $V_x <V_c$  and (b) $V_x >V_c$. 
For illustrative purposes, we use $\Delta=1$~meV in panels (a) and (b). For additional information,
see Supplemental Material \cite{supplemental}.
Middle panels: 
Modulus of the roots inside the unit circle vs $\omega/\Delta$ for (c) $V_x<V_c$ and (d) $V_x>V_c$. Solid curves represent numerically exact results 
and dashed curves follow from Eqs.~(\ref{ansatz}) and (\ref{ansatz2}), respectively. 
Bottom panels: Energy dependence of the  local bulk DoS, $\rho(\omega)$ (in meV$^{-1}$),  for (e) $V_x<V_c$ and (f) $V_x>V_c$. 
The solid red curves depict numerically exact results using Eq.~\eqref{GR11} and the dashed green curves show
approximate results obtained from Eq.~(\ref{approximate-GF}). }
\label{fig2}
\end{figure} 

Next we turn to the topologically nontrivial regime $V_x >V_c$, where the momentum $k_1$ in Eq.~\eqref{k12} becomes purely imaginary.
We should then replace $\delta_1\to i\delta_1$ in the above ansatz for the roots.
As a consequence, the $z_{\nu=1}(\omega)$ roots become real-valued, and the ansatz for  $V_x>V_c$ takes the form 
\begin{eqnarray}\label{ansatz2}
z_{1,\pm}(\omega) &=& \left(1\pm\tau_1\sqrt{\Delta_1^2 - \omega^2}\right)\ e^{- \delta_1}, \nonumber \\
z_{2,\pm}(\omega) &=& \left(1-\tau_2\sqrt{\Delta_2^2 - \omega^2}\right) \ e^{\pm i \delta_2} ,
\end{eqnarray} 
where both $\delta_1$ and $\delta_2$ are real positive.  We thus have only a single pair of complex conjugate roots ($z_2$) near the unit circle for $V_x>V_c$. 
Accurate analytical results for the  $\delta_\nu$ and $\tau_\nu$ parameters
can be obtained by solving a cubic equation (see Appendix~\ref{appA}).
As illustrated in Fig.~\ref{fig2}(c) and \ref{fig2}(d), Eq.~\eqref{ansatz2} captures the low-energy 
behavior of the roots rather well, especially in cases where electron-hole symmetry is approximately realized.

For this model of symmetry class D, the $\mathbb{Z}_2$ bulk topological invariant takes the form \cite{Alicea2012}
\begin{equation}\label{bulkinv}
Q = \frac{\text{sgn} \, \text{Pf}\, \hat{\mathcal{H}}(k=0)}{\text{sgn} \, \text{Pf}\, \hat{\mathcal{H}}(k=\pi/a)} = \pm 1.
\end{equation} 
Interestingly, the number $N_p$ of complex conjugate root pairs near (but inside) the unit circle is in correspondence with the 
topological invariant, $Q=(-1)^{N_p}$. These roots can be
unambiguously identified as the ones approaching the unit 
circle from inside in the limit $\Delta \rightarrow 0$, corresponding to the Fermi points in the normal phase. 
For an odd (even) number of pairs, the phase is thus topogically nontrivial with $Q=-1$ (trivial with $Q=1$). The upper panels in Fig.~\ref{fig2} illustrate the distribution of the roots inside the unit circle for the cases $V_x<V_c$ and $V_x>V_c$.
We observe that upon entering the topologically nontrivial regime, the complex conjugate $z_1$ roots coalesce to form an almost 
degenerate root pair $z_{1,\pm}$ [see Eq.~\eqref{ansatz2}] located on the real axis inside the unit circle.  The roots on the real axis correspond to additional bands at high 
energies above $\Delta$.  At the same time, a single pair of complex conjugate roots ($z_2$) remains near (but inside) the unit circle, as one expects for a topologically nontrivial phase.
As remarked above, this change in the structure of the roots across the transition is consistent with the corresponding change in the topological invariant. 
The transition between both regions happens when the Pfaffian, or equivalently the Hamiltonian determinant, at $k=0$ vanishes. Using the relation 
$\det\hat{\mathcal{H}}(k=0) =\prod_{n=1}^8 (1-z_n(0))$, we thus reproduce property (iv) in Sec.~\ref{sec2} which signals the phase transition. 
It is also worth mentioning that the bulk invariant \eqref{bulkinv} can be directly expressed in terms of bGFs for the semi-infinite wire:
Using $Q = \det\hat{r}_L =  \det\hat{r}_R$ (see Ref.~\cite{Fulga2012}) the reflection matrices $\hat r_{L/R}$ and therefore also $Q$
can be obtained from the bGFs [see Eq.~(\ref{reflection_matrix})].

The knowledge of the roots also gives access to other electronic properties of interest. For instance, we can obtain a compact expression for the energy-dependent local
DoS at, say, lattice site $j=0$ of the translation-invariant chain: 
\begin{equation}\label{bulkdos}
\rho(\omega)=-\frac{1}{\pi} {\rm Im}\, {\rm Tr}\left[ \hat G_{00}(\omega)\right].
\end{equation}
We focus on the low-energy limit, where one can expand the cofactor matrix $\hat A(z,\omega)$ in Eq.~\eqref{cofactor} to linear order in $\omega$. 
The local GF then follows as
\begin{equation}\label{GF00}
\hat{G}_{00}(\omega)  \simeq \sum_{|z_n|<1} \frac{\hat{A}(z_n,\omega) + \omega\hat{A}'(z_n,\omega)}{\prod_{m\ne n}\left(z_n - z_m\right)} ,
\end{equation}
where $\hat{A}'(z_n,\omega)=\frac{d}{d\omega}\hat{A}(z_n(\omega),\omega)$.  
From our ansatz in Eqs.~\eqref{ansatz} and \eqref{ansatz2},  the sum in Eq.~\eqref{GF00} can be reconstructed. 
A simple approximate expression follows for small $\Delta$ in the low-energy limit, 
where one needs to keep just the first-order terms $\propto \tau_\nu\sqrt{\Delta_\nu^2 - \omega^2}$ in the denominator. We then obtain 
\begin{equation}\label{approximate-GF}
\hat{G}_{00}(\omega) \approx \sum_{\nu=1,2} \frac{\hat{A}_{\nu} + \omega \hat{A}'_{\nu}}{\sqrt{\Delta^2_{\nu} - \omega^2}},
\end{equation} 
where $\hat{A}_\nu$ and $\hat{A}'_\nu$ are specified in Appendix~\ref{appA}.
We note that for $V_x>V_c$, the main contribution to Eq.~\eqref{approximate-GF} stems from the residues associated to $z_2$.
The results for $\rho(\omega)$ depicted in Figs.~\ref{fig2}(e) and \ref{fig2}(f) demonstrate that 
Eq.~\eqref{approximate-GF} accurately reproduces numerically exact calculations, both below and above the topological transition. 

\begin{figure}
\includegraphics[width=\columnwidth]{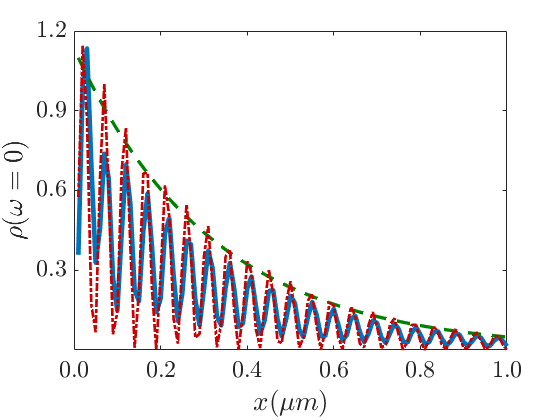}
\caption{Spatial variation of the local DoS, $\rho_j(0)$ (in meV$^{-1}$), vs distance from the boundary, $x=ja$ (in $\mu$m), for the $\omega=0$ Majorana state 
in a semi-infinite TS wire with $\mu = 1$~meV and $V_x = 2 V_c$.  
The solid blue curve gives numerically exact results obtained from Eq.~\eqref{Dyson-n}. Red-dotted  and green-dashed curves show  Eq.~(\ref{decay}) with and without $2k_F$ oscillations, respectively. }
\label{fig3}
\end{figure} 

Next we turn to the case of a semi-infinite chain in the topological phase, $V_x>V_c$.
Using the Dyson equation in Eq.~(\ref{Dyson-n}) and taking into account the behavior of the roots
of the infinite chain discussed above, we can deduce the spatial decay profile of the zero-energy Majorana end state into the bulk. Noting    
 that the GF components $\hat G_{j,0}$ and $\hat G_{0,j}$ in Eq.~(\ref{Dyson-n}) are $\propto |z|^{j}$, we observe that for $V_x > V_c$, the decay is dominated by the $z_2$ roots since $|z_2|>|z_1|$. Moreover, the
 decay profile  exhibits fast oscillations due to the complex phase $\delta_2$ in Eq.~\eqref{ansatz2},
which for $\mu \gg \Delta$ can be approximated as $\delta_2\simeq k_F a$ with $k_F\equiv k_2$.  In this approximation, the local DoS of the $\omega=0$ MBS thus has the spatial profile 
\begin{equation}\label{decay}
\rho_j(\omega=0) \propto |z_2(0)|^{2j} \cos^2\left(jk_Fa + \chi_0 \right) ,
\end{equation}
where $\chi_0$ describes a phase shift in the $2k_F$ oscillations.   Equation~(\ref{decay})  reproduces 
the numerically exact results obtained from Eq.~\eqref{Dyson-n} rather well, as illustrated in Fig.~\ref{fig3}.  The dashed curve shows that the envelope function is accurately described by $|z_2(0)|^{2j}$, corresponding to an exponential decay into the bulk of the chain.

\section{Two-channel class-D nanowire}\label{sec4}

We next examine the case of spinful multichannel hybrid nanowires with broken time reversal symmetry.  The bGF approach could in principle be applied to nanowire 
models with an arbitrary number of channels. In practice, however, the techniques in Sec.~\ref{sec2} are less efficient once the degree $2N$ of the secular polynomial 
\eqref{polysec} becomes very large. We here restrict ourselves to the two-channel case with $N=8$, which can be realized for two single-channel nanowires coupled by tunneling terms.
The resulting model already exhibits many of the features expected for generic multichannel nanowires \cite{Lutchyn2011a,Woods2019}. 

Our model Hamiltonian is given by
\begin{equation}\label{H2ch}
\hat{\mathcal{H}}_{\rm 2ch}(k) = \left(\begin{array}{cc} \hat{\mathcal{H}}(k) & \hat{T} \\
\hat{T}^{\dagger} & \hat{\mathcal{H}}(k) \end{array}\right),
\end{equation}
where the $2\times 2$ structure refers to  wire space.  We consider two identical spinful single-channel Majorana wires described by the model of Refs.~\cite{Lutchyn2010,Oreg2010} with $\hat{\mathcal{H}}(k)$ in Eq.~\eqref{Lutchyn}. The interwire tunnel couplings are modeled by
\begin{equation}\label{interwire}
\hat{T} =-t_y \sigma_0\tau_z + i \alpha_y \sigma_x \tau_z  + \Delta_y \sigma_0\tau_x,
\end{equation}
where $t_y$  and $\alpha_y$ are spin-conserving and spin-flipping hopping amplitudes, respectively. 
The coupling $\alpha_y$ may arise due the presence of a Rashba spin-orbit coupling produced by an electric field along the $z$-direction.
As in Sec.~\ref{sec3}, we write $t_y= \hbar^2/(2m^* a_y^2)$ and $\alpha_y = \hbar u/a_y$, with the minimal distance $a_y$ between the wires. 
In the concrete examples shown below, we assume $a_y=3 a$, which corresponds to a subband separation of $\approx 3$~meV.  
The interwire coupling \eqref{interwire} also includes a non-local inter-wire pairing amplitude $\Delta_y$. For the present class D case, however, we find that
allowing for a small $\Delta_y\ne 0$ does not lead to significant changes in the phase diagram.  We thus put $\Delta_y=0$ in this section.

\begin{figure*}[t]
\includegraphics[width=\textwidth]{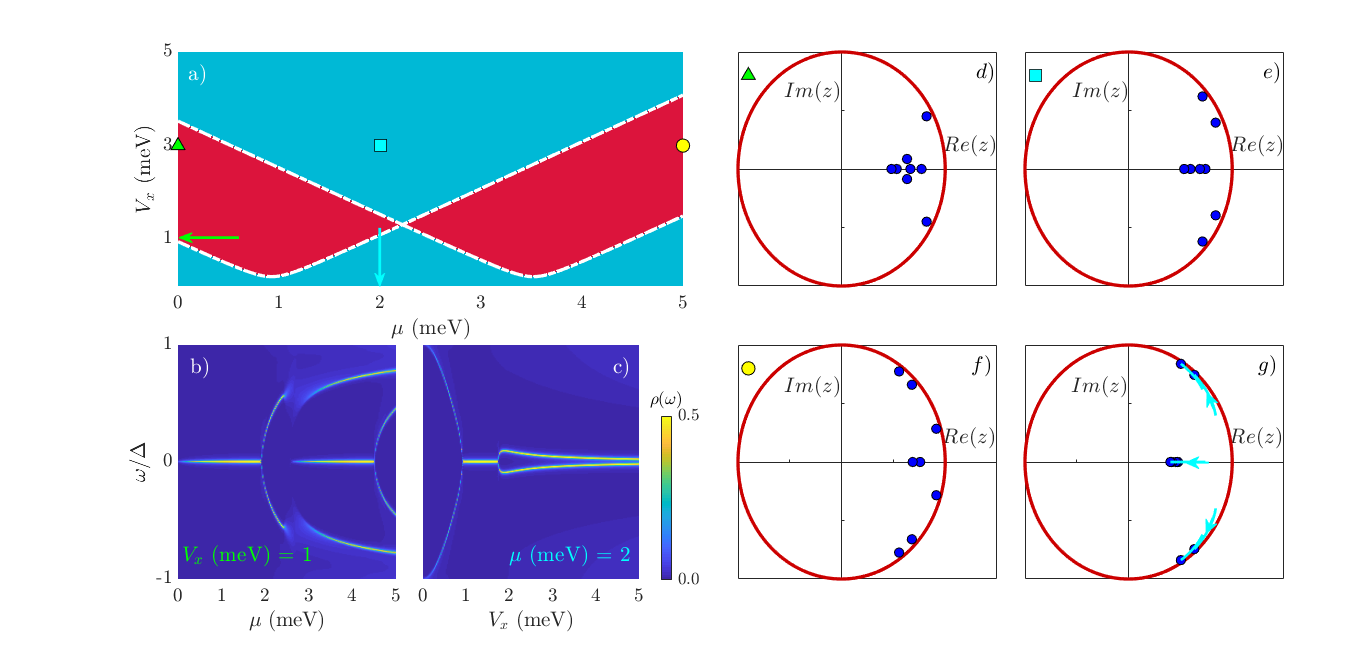}
\caption{Two-channel spinful Majorana wire model of class D [see Eq.~\eqref{H2ch}] with parameters as explained in the main text.
Panel (a) shows the bulk phase diagram in the $\mu$-$V_x$ plane. Topological nontrivial (trivial) phases are shown in red (blue). Panels (b) and (c) show the energy dependence of the local DoS, 
$\rho_{j=1}(\omega)$ (in meV$^{-1}$), at the boundary of a semi-infinite two-channel wire along the trajectories marked by arrows in panel (a). 
Panels (d)--(f) illustrate the roots $z_n(0)$ inside the unit circle at the three points indicated in panel (a) by a triangle (d), a square (e), and a circle (f), respectively. 
For additional insights,
see Supplemental Material \cite{supplemental}. Panel (g) shows the evolution of the roots within the topologically trivial regime as $V_x$ increases  from 3 to 8 meV at constant chemical potential $\mu=2$~meV. In panels (d)--(g), we use $\Delta=1$~meV.}
\label{fig4}
\end{figure*}
 
One can characterize the phase diagram of a translationally invariant two-channel wire by using the bulk topological invariant in Eq.~\eqref{bulkinv}
with the replacement $\hat{\cal H}(k)\to \hat{\cal H}_{\rm 2ch}(k)$. The Pfaffian at $k=0$ is here given by 
\begin{eqnarray}
\mbox{Pf}\,\hat{\mathcal{H}}_{\rm{2ch}}(0) &=& \alpha_y^4+\left[(\mu-3t_y)^2-V_x^2+\Delta^2\right] \nonumber \\
 &\times&\left [(\mu-t_y)^2-V_x^2+\Delta^2\right] \nonumber \\
&+&2\alpha_y^2\left[-(\mu-3t_y)(\mu-t_y)-V_x^2+\Delta^2\right]. \quad
\label{Pf2ch}
\end{eqnarray}
The boundaries of the topological phase correspond to a vanishing Pfaffian at $k=0$, where Eq.~\eqref{Pf2ch} implies the two critical Zeeman fields
\begin{eqnarray}\nonumber
V_{c,\pm} &=& \Bigl( \alpha_y^2+\mu^2-4\mu t_y+5t_y^2+\Delta^2  
\\ &&\quad \pm\   2|\mu-2t_y|\sqrt{t_y^2+\alpha_y^2} \Bigr)^{1/2}.\label{borderTP}
\end{eqnarray}
The resulting phase diagram in the $\mu$-$V_x$ plane is illustrated in Fig.~\ref{fig4}(a).  We observe that
the two-channel model \eqref{H2ch} exhibits a richer phase diagram than in the single-channel case (see also  Refs.~\cite{Lutchyn2011a,Woods2019}).

We next construct the bGF of a semi-infinite wire by determining the roots of the secular polynomial in Eq.~\eqref{polysec}, which here is a 16th-order polynomial equation 
that we solve numerically. Figure \ref{fig4}(b) and \ref{fig4}(c) illustrates the evolution of the energy-dependent local DoS, $\rho_1(\omega)$, at the boundary, i.e., taken at site $j=1$ of a
semi-infinite two-channel wire.  We consider two different trajectories in the $\mu$-$V_x$ plane as indicated by the arrows in Fig.~\ref{fig4}(a). For constant $V_x$ [panel b)], 
there are both topologically nontrivial and trivial regions as $\mu$ is varied. In 
the topologically nontrivial regions, we observe a zero-energy peak in the local DoS, signaling the presence of MBSs.  
This $\omega=0$ peak is absent in the trivial regime. For fixed $\mu$ [panel (c)], the topologically nontrivial phase is reached for intermediate values of $V_x$. For larger $V_x$, even though the system is in a trivial phase,
 we find low-energy Andreev bound states that approach zero energy as $V_x$ increases. This effect has also been described in Ref.~\cite{Woods2019}. 

Additional insights follow by analyzing the evolution of the roots $z_n(\omega=0)$ inside the unit circle
in the complex momentum plane. In Fig.~\ref{fig4}(d)--(f), we illustrate their distribution for three different points in the phase diagram. 
For panels (d) and (f), the system is in a topological phase and, as expected, one finds an \emph{odd} number of pairs of complex conjugate roots close to the unit circle.
As in Sec.~\ref{sec3}, the roots on the real axis correspond to additional bands at higher energies well above $\Delta$. Panel (e) instead corresponds to a topologically trivial phase with an 
\emph{even} number of conjugate  root pairs near the unit circle.
Finally, Fig.~\ref{fig4}(g) illustrates the evolution of the roots in the topologically trivial regime as the Zeeman parameter $V_x$ increases. We find 
that the both roots near the unit circle in the first quadrant become almost degenerate for large $V_x$.  Such a behavior effectively amounts to having two replicas of a single-channel TS wire, which in turn helps to explain why Andreev bound states approach the zero energy limit for strong Zeeman field [see Fig.~\ref{fig4}(c) and Ref.~\cite{Woods2019}].  

\section{TRITOPS nanowires}
\label{sec5}

Next we turn to models for hybrid nanowires of symmetry class DIII. Such TRITOPS wires constitute another interesting system with topologically nontrivial phases.
Below we first study single-channel wires and subsequently turn to the two-channel case. 

\subsection{Single-channel case}\label{sec5a}

Many different proposals for physical realizations of single-channel TRITOPS wires have been put forward in the recent past \cite{Wong2012,Nakosai2012,Deng2012,Zhang2013,Keselman2013,Dumitrescu2013,Chung2013,Nakosai2013,Haim2014,Gaud2014,Klinovaja2014,Schrade2015,Haim2016,Kim2016,Mellars2016,Camjayi2017,Schrade2018,Haim2019a}.  For concreteness, we will here focus on the  model introduced by Zhang \emph{et al.}~\cite{Zhang2013}. Using the spin-Nambu basis with $N=4$ in Sec.~\ref{sec2}, the Hamiltonian is given by
\begin{equation}\label{DIII-single}
\hat{\mathcal{H}}_{\rm DIII}(k) = \epsilon_k \sigma_0\tau_z  + \alpha_k\sigma_z\tau_z + \Delta_k\sigma_0\tau_x,
\end{equation}
where in this section, we use 
\begin{eqnarray}\nonumber
    \epsilon_k &=& -2t\cos(ka)-\mu,\quad \alpha_k=2\alpha\sin(ka),\\ &&\qquad \Delta_k=2 \Delta \cos(ka).
\end{eqnarray}
Again $t$ corresponds to a nearest-neighbor hopping amplitude, $\mu$ is the chemical potential, $a$ the lattice spacing, and $\alpha$ the spin-orbit coupling strength. The 
parameter $\Delta$ corresponds to a nearest-neighbor pairing interaction.  In the examples below, we use $a=10$~nm, $t=10$~meV and $\alpha=4$~meV as in Secs.~\ref{sec3} and \ref{sec4}. 

By a simple rearrangement of the spin-Nambu spinor $\hat \Psi_k$, one can block-diagonalize the Hamiltonian in Eq.~\eqref{DIII-single},
$\hat{\mathcal{H}}_{\rm DIII} = {\rm diag}(\hat{\mathcal{H}}_-, \hat{\mathcal{H}}_+)$.  To that end, upon replacing $\hat{\Psi}_k^T\to
\left(c_{k\uparrow}, c^{\dagger}_{-k\downarrow},c_{k\downarrow},-c^{\dagger}_{-k\uparrow}\right)$, we arrive at the $2\times 2$ block 
Hamiltonians
\begin{equation} \label{kane_TRITOPS}
\hat{\mathcal{H}}_\pm(k) = (\epsilon_k \mp \alpha_k)\tilde{\sigma}_z  + \Delta_k \tilde{\sigma}_x = \beta_\pm(k) \cdot \tilde{\sigma},
\end{equation}
where $\tilde{\sigma}$ is the vector of $\tilde{\sigma}_{x,y,z}$ Pauli matrices 
in the respective $2\times 2$ space obtained after block diagonalization.
Each Hamiltonian $\hat {\cal H}_\pm(k)$ corresponds to a Dirac-type model where
\begin{equation}\label{beta}
\beta_\pm(k) = \left(\begin{array}{c} 2\Delta\cos(ka)\\ 0\\     -\mu-2t\cos(ka)\pm 2\alpha\sin(ka)\end{array}\right)
\end{equation}
is a vector field mapping the first Brillouin zone onto a closed curve.

\begin{figure}[t]
\includegraphics[width=0.8\columnwidth]{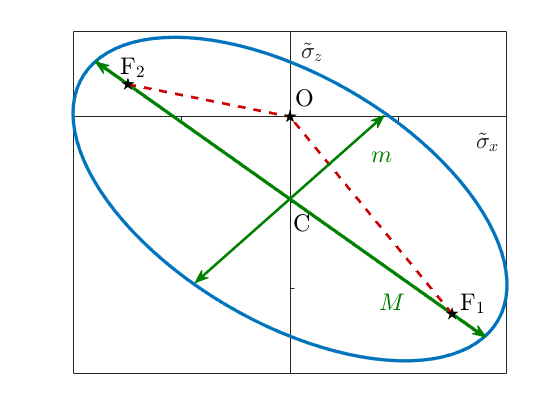}
\caption{Curve traced out by $\beta_-(k)$ in the $\tilde{\sigma}_x$-$\tilde{\sigma}_z$ plane for a single-channel TRITOPS wire in a topologically nontrivial phase [see Eqs.~\eqref{kane_TRITOPS} and \eqref{beta}] with
$t = 0.5, \alpha = 0.8, \Delta=1,$ and $\mu=1.04$ (all in meV). The evolution of the 
bulk Hamiltonian $\hat{\cal H}_-(k)$ upon traversal of the Brillouin zone is described by an
ellipse containing the origin (O). For details, see main text and Appendix~\ref{appB}.}
\label{fig5}
\end{figure}

At this stage, we can apply the formalism of Ref.~\cite{Mong2011} for analyzing the roots of the secular polynomial of Dirac-like Hamiltonians. 
By projecting $\hat{\mathcal{H}}_{\pm}$ to the $\tilde{\sigma}_x$-$\tilde{\sigma}_z$ plane, we obtain an elliptic curve as illustrated in Fig.~\ref{fig5}. According to the arguments in Ref.~\cite{Mong2011}, if the ellipse encloses the origin of the $\tilde{\sigma}_x$-$\tilde{\sigma}_z$ plane, we know that for a semi-infinite wire, 
$\hat{\cal H}_\pm(k)$ will generate an edge state with energy equal to the modulus of the component of $\beta_\pm(k)$ 
perpendicular to this plane.
In our case,  $\left[\beta_{\pm}(k)\right]_y=0$ implies that 
we have a pair of zero-energy boundary states in the topological phase. 
In addition, this argument also shows that there are no finite-energy Andreev bound states in the trivial phase (where the ellipse does not contain the origin).
For the case in Fig.~\ref{fig5}, where the origin is displaced along the $\tilde{\sigma}_z$-axis, the topological transition occurs at $ka=\pm \pi/2$ and $|\mu| = 2\alpha$ [see Eq.~\eqref{beta}]. 
This conclusion is consistent with the fact that at the topological transition, one finds roots at 
$z=e^{ika}=\pm i$ (see also Ref.~\cite{Aligia2018}), in agreement with property (v) in Sec.~\ref{sec2}.

\begin{figure}
\includegraphics[width=\columnwidth]{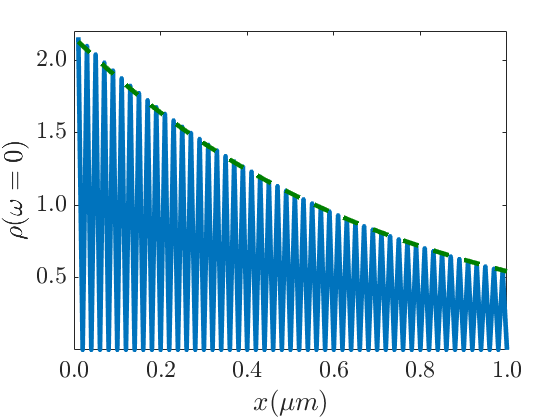}
\caption{Spatial variation of the local DoS at zero energy (in meV$^{-1}$), corresponding to Majorana end states of a semi-infinite TRITOPS wire in its topological phase [see
Eq.~(\ref{DIII-single})] for $\mu=0$ (blue solid curve). 
The green dashed curve shows an exponential decay on the length scale $\lambda_e= -\frac{a}{2}\ln|z_{\rm max}|$ [see Eq.~(\ref{zmax-tritop})].}
\label{fig6}
\end{figure}

More generally, by determining the roots $z_n(\omega)$, we can again construct the bGF of a semi-infinite wire. In particular, we thereby obtain the class DIII bulk 
topological invariant via the reflection matrices in Eq.~(\ref{reflection_matrix}). 
In the present case, the invariant is given by
$Q = {\rm Pf}\left(i\hat{r}_{L,R}\right)$ \cite{Fulga2012}.  Furthermore,
using the results of Refs.~\cite{Mong2011,Aligia2018}, an analytical expression for the largest-modulus zero-frequency root, $z_{\rm max}$, inside the unit circle can be computed from purely geometrical considerations for the ellipse in Fig.~\ref{fig5} (see Appendix~\ref{appB} for details).
The length scale governing the spatial decay profile of the pair of Majorana states localized near the boundary of a semi-infinite TRITOPS wire then follows as $\lambda_e = -\frac{a}{2}
\ln|z_{\rm max}|$ [see Eq.~\eqref{zmax-tritop} in Appendix~\ref{appB}].
The validity of this expression is confirmed in Fig.~\ref{fig6}, where we show numerically exact results for the spatial variation of the local DoS at $\omega=0$ together with the prediction obtained from Eq.~(\ref{zmax-tritop}).

\subsection{Two-channel case}\label{sec5b}

As in Sec.~\ref{sec4}, we can also extend the TRITOPS model to the two-channel case by coupling two single-channel wires.
More general multichannel wire constructions are also possible but will not be pursued here.
The corresponding Hamiltonian is with Eq.~\eqref{DIII-single} given by
\begin{equation}
\hat{\mathcal{H}}_{\rm DIII, 2ch}(k) = \left(\begin{array}{cc} \hat{\mathcal{H}}_{\rm DIII}(k) & \hat{T}_{\rm DIII} \\
\hat{T}_{\rm DIII}^{\dagger} & \hat{\mathcal{H}}_{\rm DIII}(k) \end{array}\right),
\label{multichannel-DIII}
\end{equation}
where the interwire tunneling couplings are modeled in a similar manner as in Eq.~\eqref{interwire}:
\begin{equation}
    \hat{T}_{\rm DIII} =-t_y \sigma_0\tau_z + i \alpha_y \sigma_y \tau_z+\Delta_y \sigma_0\tau_z.
\end{equation}
We here allow for
spin-conserving $(t_y)$ and spin-flipping $(\alpha_y)$ hopping processes, as well as for nonlocal pairing terms ($\Delta_y$). Below, $t_y$ and $\alpha_y$ are parametrized as specified in Sec.~\ref{sec4}.

\begin{figure}[t]
 \centering 
  \includegraphics[width=1\columnwidth]{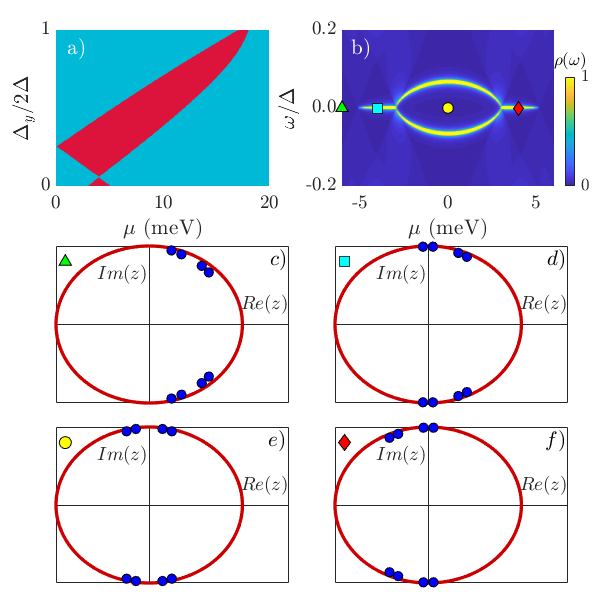}
\caption{Two-channel TRITOPS nanowire [see Eq.~\eqref{multichannel-DIII}], with parameters as explained in the main text. Panel (a) shows the phase diagram in the $\mu$--$\Delta_y$ plane, with the topologically nontrivial (trivial) phase  in red (blue). (b) Local DoS, $\rho_{j=1}(\omega)$ (in meV$^{-1}$), at the boundary of a semi-infinite wire
in the $\mu$--$\omega$ plane for $\Delta_y=0$. Panels (c) to (f) depict the roots $z_n(\omega=0)$ inside the unit circle for different $\mu$ as indicated by the respective symbol in panel (b).
We use $\Delta=1$~meV in panels (c)--(f).}
\label{fig7}
\end{figure}

The resulting phase diagram is illustrated in Fig.~\ref{fig7}(a).  To make analytical progress, from now on we consider the case $\Delta_y=0$
and determine the conditions for gap closings, and thus for phase transition curves in the two-channel TRITOPS case.  
The gap closes again for $ka = \pm \pi/2$ as in Sec.~\ref{sec5a} but now for the chemical potential set to one of the critical values
\begin{equation}
|\mu_{\pm}| = \sqrt{\alpha_y^2 + (t_y \pm 2\alpha)^2}.
\end{equation}
 where the topological invariant is 
related to the product of the signs of the effective pairing amplitude at different Fermi points \cite{Zhang2010}. 
As the critical momenta are as in Sec.~\ref{sec5a}, the pairing function is directly determined by $\Delta \cos(ka)$ [see Eq.~(\ref{DIII-single})]. 
For this reason, the topologically nontrivial (trivial) phase has an odd (even) number of Fermi points between $ka=0$ and $ka=\pi/2$.

The bGF can again be computed from the roots of the secular polynomial. The latter also determine the behavior of the edge modes of a semi-infinite
two-channel TRITOPS wire in different regions of the phase diagram. 
By continuity, the condition of having an odd number of Fermi points with $0<k_F<\pi/2a$ corresponds to an odd number $N_p$ of roots near the unit circle in the first quadrant.
Our results for the roots are illustrated in Fig.~\ref{fig7}(c)--\ref{fig7}(f).  As expected,  $N_p$ is odd for panels (d) and (f), where panel (b) shows that Majorana end states are present and thus a topological phase is realized.  By contrast, panels (c) and (e) show topologically trivial cases with even $N_p$.

\section{Phase-biased topological Josephson junctions}\label{sec6}

In this section, we consider different examples for the equilibrium supercurrent-phase relation in two- and three-terminal Josephson junctions made of nanowires in topologically nontrivial superconducting phases.
These wires are coupled together by tunnel junctions.  We start in Sec.~\ref{sec6a} with the case of a trijunction of TS nanowires
(see also Ref.~\cite{Zazunov2017}), and then turn to TRITOPS-TS Josephson junctions in Sec.~\ref{sec6b}.

\subsection{Three-terminal TS junctions} \label{sec6a}

\begin{figure}
\includegraphics[width=\columnwidth]{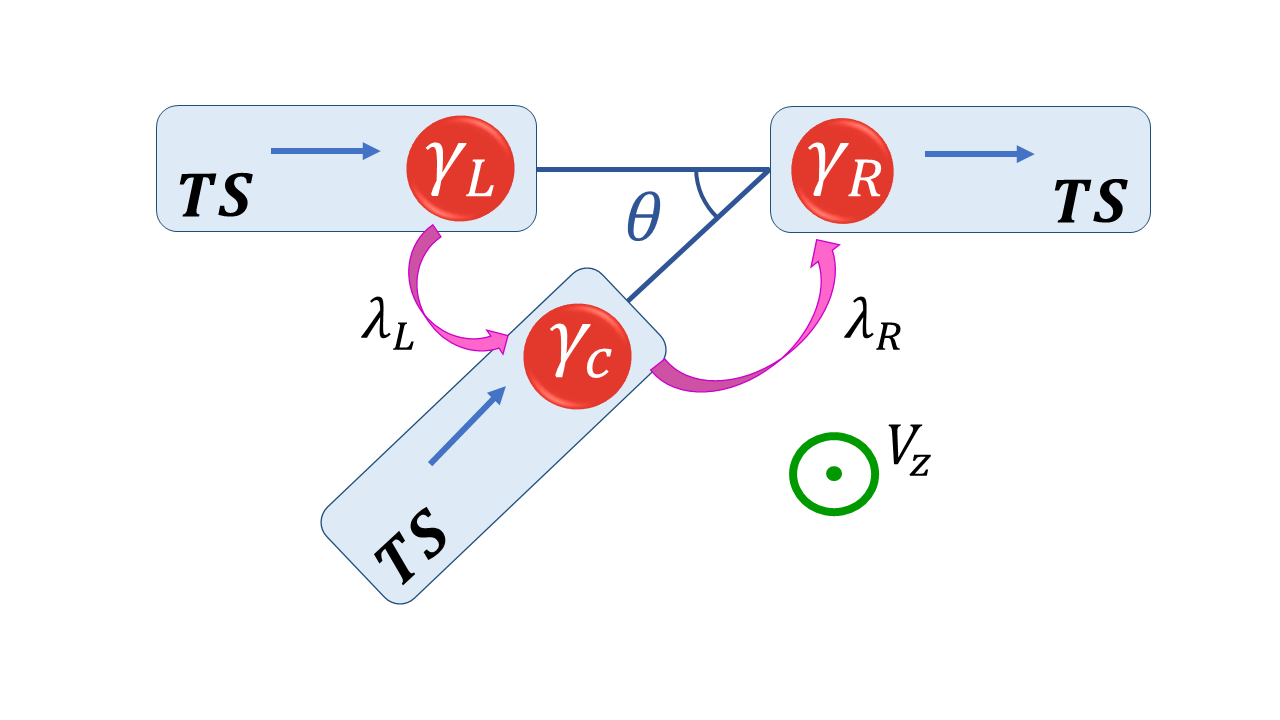}
\caption{Three-terminal junction of spinful TS nanowires (see Sec.~\ref{sec3}), with two parallel wires ($L,R$) and a central $(C)$ wire at angle $\theta$.  
The red dots indicate MBSs with Majorana operators $\gamma_{L,R,C}$ near the junction, with tunnel couplings $\lambda_{L,R}$
connecting the $L,R$ wires to the $C$ wire.  
We assume that no direct tunnel coupling between the $L$ and $R$ wires is present.
A Zeeman field $V_z$ is applied perpendicular to the plane containing the three wires.
Blue arrows show the positive momentum direction in each wire.}
\label{fig8}
\end{figure}

We first consider a three-terminal junction formed by spinful single-channel nanowires in the TS phase. For a schematic layout, see Fig.~\ref{fig8}.  Such devices have been
suggested, e.g., for Majorana braiding implementations \cite{Alicea2011,Deb2018,Trif2019},  for the engineering of artificial topological Weyl semimetal phases
\cite{Gavensky2019,Meyer2020}, and for the observation of giant shot noise features induced by the single zero-energy  MBS 
localized at the trijunction  \cite{Jonckheere2019}.
While most previous studies have been based on minimal models or on spinless Kitaev chain models, a more realistic description using the spinful nanowire model of
Refs.~\cite{Lutchyn2010,Oreg2010} discussed in Sec.~\ref{sec3} is desirable. In particular, one can then assess the role of the spin degree of freedom
and the effects of various microscopic parameters such as the angle $\theta$ in Fig.~\ref{fig8}. We assume that each wire is sufficiently long such that the overlap between MBSs located at different ends of the same wire is negligibly small. 

We model each nanowire in the setup of Fig.~\ref{fig8} in terms of the spinful single-channel Hamiltonian of Eq.~(\ref{Lutchyn}).  All three wires lie in a plane, with two of them aligned ($L$ and $R$ in Fig.~\ref{fig8}) and the third (the central wire, $C$, in Fig.~\ref{fig8}) at an arbitrary angle $\theta$ to the other two. 
We here assume that the Zeeman field $V_z$ is oriented perpendicular to the plane (see Ref.~\cite{Alicea2012}). 
For simplicity, we  consider identical material parameters for the three wires which are chosen such that the TS phase is realized.

Let us next discuss the unitary rotations necessary to adapt the bGFs of Sec.~\ref{sec3} to a common reference frame for all three wires in Fig.~\ref{fig8}.
We first perform a $\pi/2$ rotation of the spin axis around the $y$ axis, which connects the intrinsic coordinate system of the $L$ and $R$ wires
to the common reference frame. Defining
\be\label{rtheta}
R(\vartheta) = \left[\sigma_0 \cos(\vartheta/2)-i \sigma_y \sin(\vartheta/2)\right]\tau_0 ,
\ee
the corresponding rotation matrix, $R_y=R(\vartheta=\pi/2)$, transforms a Zeeman field along the $x$-direction (see Sec.~\ref{sec3}) into a Zeeman field along
the negative $z$-direction (as in Fig.~\ref{fig8}). The  bGFs for the $L$ and $R$ wires in Fig.~\ref{fig8} are thus given by 
\be 
\hat{\cal G}_{L/R}'= R_y \hat{\cal G}_{L,R}^{} R_y^{-1},
\ee
with $\hat {\cal G}_{L,R}$ as described in Sec.~\ref{sec3}.  
For the $C$ lead, we  additionally have to rotate by the angle $\theta$ around the global $z$-axis. The corresponding rotation matrix, $R_z(\theta)$, follows from Eq.~\eqref{rtheta} with the replacements $\sigma_y\to \sigma_z$ and $\vartheta\to\theta$. We thereby obtain 
\be 
\hat{\cal G}_C = R_{z}(\theta) R_y \hat{\cal G}_L R_y^{-1} R^{-1}_{z}(\theta).
\ee
In what follows, we rewrite $\hat{\cal G}_{L/R}'\to \hat{\cal G}_{L/R}$ to keep the notation simple. 

\begin{figure}[t]
 \centering
  \includegraphics[width=\columnwidth]{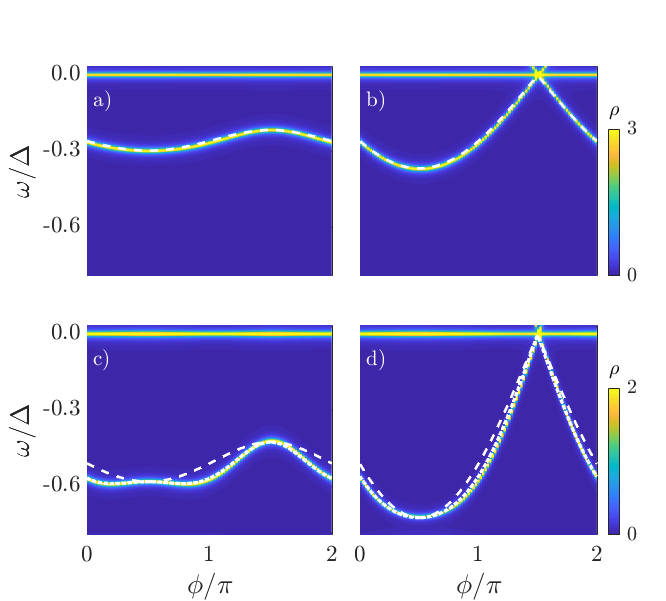}
\caption{Phase dependence of the subgap spectrum of the trijunction of TS wires in Fig.~\ref{fig8}, with 
the superconducting phases $\phi_L=-\phi_R=\phi$ and $\phi_C=0$.
The TS wires are modeled as spinful nanowires with $\mu = 2$~meV, $V_z = 3V_c$, and symmetric couplings, $\lambda_L=\lambda_R=\lambda$. For other parameters, see Sec.~\ref{sec3}.
Panel (a) [(b)] is for $\lambda=2$~meV and $\theta=\pi/2$ [$\theta=\pi/10$].
Panel (c) [(d)] is for $\lambda=5$~meV and $\theta=\pi/2$ [$\theta=\pi/10$].
From blue to yellow,  $\rho_{\rm 3TS}(\omega)$ (in meV$^{-1}$) gradually increases, where Eq.~\eqref{3ldos} 
has been evaluated in a numerically exact manner.
White dotted [dashed] curves show the approximate Andreev bound state dispersion relation in Eq.~\eqref{abs2} [Eq.~\eqref{abs1}].}
\label{fig9}
\end{figure}

The coupling between the $L,R$ wires and the $C$ wire is modeled by a spin-conserving tunneling term,
\be \label{dyson-trijunction}
H_T = \frac{1}{2}\sum_{\nu = L,R}\hat{\Psi}^\dag_\nu\ \hat\lambda_\nu \hat{\Psi}_C^{} + {\rm H.c.},
\quad \hat{\lambda}_{\nu} = \lambda_\nu \sigma_0\tau_z e^{i\tau_z\phi_\nu/2},
\ee 
where $\hat{\Psi}_{L,R,C}$ are boundary spinor fields and $\phi_\nu$ is the phase of the superconducting order parameter 
in the respective wire. We choose a gauge with $\phi_C=0$ and real-valued tunnel couplings $\lambda_\nu$.
The physical properties of the trijunction are then determined by the full bGF,
\be\label{G3TS}
\hat{G}_{\rm 3TS} = \left(\begin{array}{ccc} \hat{\cal G}_L^{-1} & \hat{\lambda}_L & 0 \\
                                                   \hat{\lambda}_L^\dagger & \hat{\cal G}_C^{-1} & \hat{\lambda}_R \\
                                                   0 & \hat{\lambda}_R^\dagger & \hat{\cal G}_R^{-1} \end{array}\right)^{-1} ,
\ee 
where the $3\times 3$ structure refers to wire space. From Eq.~\eqref{G3TS}, the energy dependence of the local DoS at the junction will be given by
\begin{equation}\label{3ldos}
    \rho_{\rm 3TS}(\omega)=-\frac{1}{\pi}\mbox{Im}\, \mbox{Tr} \left[ \hat{G}_{\rm 3TS}(\omega)\right].
\end{equation}
Figure~\ref{fig9} shows the phase dependence of $\rho_{\rm 3TS}(\omega)$  obtained by numerical evaluation of Eqs.~\eqref{G3TS} and \eqref{3ldos} for a trijunction 
with   $\phi_L = - \phi_R = \phi$ and  $\phi_C=0$.   (This is the series configuration in the parlance of Ref.~\cite{Zazunov2017}.)

Deep in the topological regime, the low-energy properties of the trijunction are well described
by a minimal model keeping only the MBSs at the junction.  To show this from the above bGFs, we first derive an effective Hamiltonian for 
each wire that only keeps track of the respective MBS:
\be\label{heffnu}
H_{{\rm eff},\nu}=\lim_{\omega \to 0} \hat{\mathcal{G}}^{-1}_{\nu}(\omega). 
\ee
Using Eq.~\eqref{heffnu} and recalling that the $z_2$ roots dominate for $V_x>V_c$, we can read off the boundary spinors for each of the wires $(\nu=L,R,C$; see Ref.~\cite{Zazunov2017}):
\bea 
\hat{\Psi}_L &\simeq& \sqrt{\frac{\Delta_2}{t}} \bmat 0\\ 1 \\ -i \\ 0 \emat \gamma_L, \quad \hat{\Psi}_R \simeq \sqrt{\frac{\Delta_2}{t}} \bmat 0\\ -i \\ 1 \\ 0 \emat \gamma_R, \nonumber
\\ &&\qquad
 \hat{\Psi}_C \simeq \sqrt{\frac{\Delta_2}{t}} R_{z}(\theta)\bmat 0\\ 1 \\ -i\\0 \emat  \gamma_C,
\label{spinores}
\eea 
where the Majorana operators $\gamma_\nu$ satisfy the anticommutation relations
$\{ \gamma_\nu , \gamma_{\nu'} \} =\delta_{\nu\nu'}$. 
The pairing gap $\Delta_2$ has been defined in Sec.~\ref{sec3} [see also Fig.~\ref{fig1} and Eq.~\eqref{conditionadd}].

Next, we project the tunneling Hamiltonian \eqref{dyson-trijunction} to the Majorana sector by means of Eq.~\eqref{spinores}. We thereby arrive at a minimal model Hamiltonian,
\be\label{tunneltriminimal}
H_{\rm mm} = -i\Omega_L(\phi) \gamma_L \gamma_C-i\Omega_R(\phi)\gamma_R \gamma_C,
\ee 
with the energies
\bea\nonumber
\Omega_L(\phi) &=& \frac{2\Delta_2 \lambda_L}{t}  \sin\left(\frac{\phi + \theta}{2} \right),\\
\label{omegaLRdef}
\Omega_R(\phi) &=& -\frac{2 \Delta_2 \lambda_R}{t} \cos\left(\frac{\phi - \theta}{2}\right).
\eea
Equation \eqref{tunneltriminimal} is easily diagonalized by rotating the $\gamma_{L,R}$ operators to new Majorana operators $\tilde\gamma_{L,R}$,
\be 
\bmat \gamma_L \\ \gamma_R \emat = \bmat \sin\kappa && -\cos\kappa \\ \cos\kappa && \sin\kappa \emat \bmat \tilde{\gamma}_L \\ \tilde{\gamma}_R  \emat, 
\ee 
with $\sin\kappa = \Omega_L/\Omega$ and 
\be\label{omegadef}
\Omega(\phi) = \sqrt{\Omega_L^2(\phi) + \Omega_R^2(\phi)}.
\ee
We thereby arrive at
\be\label{mm1}
H_{\rm mm} = -i\Omega(\phi) \tilde{\gamma}_L \gamma_C,
\ee
where the decoupled Majorana operator $\tilde{\gamma}_R$ 
describes the remaining zero-energy state \cite{Jonckheere2019}. The eigenstates of Eq.~\eqref{mm1} correspond to Andreev bound states with the phase-dependent subgap energy [see Eq.~\eqref{omegaLRdef}],
\be\label{abs1}
E_\pm (\phi)=\pm \frac12 \sqrt{\Omega_L^2(\phi)+\Omega_R^2(\phi)}.
\ee
The phase derivative $\partial_\phi E_-(\phi)$ then yields the Josephson current-phase relation.
As illustrated in Fig.~\ref{fig9}, Eq.~\eqref{abs1} reproduces our numerically exact bGF calculations for small tunnel couplings $\lambda_{L,R}$. 

However, for intermediate-to-large values of the tunnel couplings, the Andreev bound state dispersion may deviate from Eq.~\eqref{abs1} [see, e.g., the `bump'-like features in Fig.~\ref{fig9}(c)].
Such deviations are due to the fact that the Majorana operators $\gamma_L$ and $\gamma_R$ will become connected through the virtual excitation of continuum quasiparticle states with above-gap energy $E>\Delta$. 
Within our minimal model, this physics can be taken into account by adding an effective coupling $\lambda_{LR}$ between the $L$ and $R$ wires. For $\lambda_\nu\ll \Delta$, we estimate
$\lambda_{LR}\simeq \lambda_L\lambda_R/\Delta$.  The corresponding tunneling term is given by
\be  
H_{T,LR} =  \frac{1}{2} \lambda_{LR}  \hat{\Psi}^\dag_L \sigma_0 e^{i\tau_z\phi}\tau_z\hat{\Psi}_R^{} + {\rm H.c.}
\ee 
Using the Majorana spinors in Eq.~(\ref{spinores}) together with Eq.~\eqref{tunneltriminimal}, we arrive at an improved version of the minimal model Hamiltonian:
\bea \label{minimal-model2} H_{\rm mm} &=&-i\Omega_L(\phi) \gamma_L\gamma_C-i\Omega_R(\phi)\gamma_R \gamma_C \nonumber\\ &&\qquad -i\Omega_{LR}(\phi) \gamma_L\gamma_R,\nonumber\\
\Omega_{LR}(\phi) &=& \frac{2 \Delta_2 \lambda_{LR}}{t} \cos \phi. 
\eea
One can easily show that Eq.~\eqref{minimal-model2} still predicts a decoupled zero-energy MBS at the trijunction. The hybridization between the remaining two Majorana states yields
Andreev bound states with the dispersion relation
\be \label{abs2}
E_\pm (\phi)= \pm \frac12 \sqrt{\Omega_L^2(\phi) + \Omega_R^2(\phi) + \Omega_{LR}^2(\phi)} .
\ee 
Of course, for $\lambda_{LR}\to 0$, we recover Eq.~\eqref{abs1}. 
Only by including the $\Omega_{LR}$ term in Eq.~\eqref{abs2}, however, the bumps found in the numerically exact dispersion in Fig.~\ref{fig9}(c) can be accurately reproduced.

We conclude that the minimal model in Eq.~\eqref{minimal-model2}, which has been derived from the bGF approach,  captures the basic
physics of the Josephson effect in the three-terminal TS junction shown in Fig.~\ref{fig8}.  In particular, the dependence of the current-phase relation
on the angle $\theta$ between the wires resulting from the subgap spectrum in Fig.~\ref{fig8} will be correctly reproduced.

\subsection{TRITOPS-TS junction} \label{sec6b}

\begin{figure}[t]
\includegraphics[width=\columnwidth]{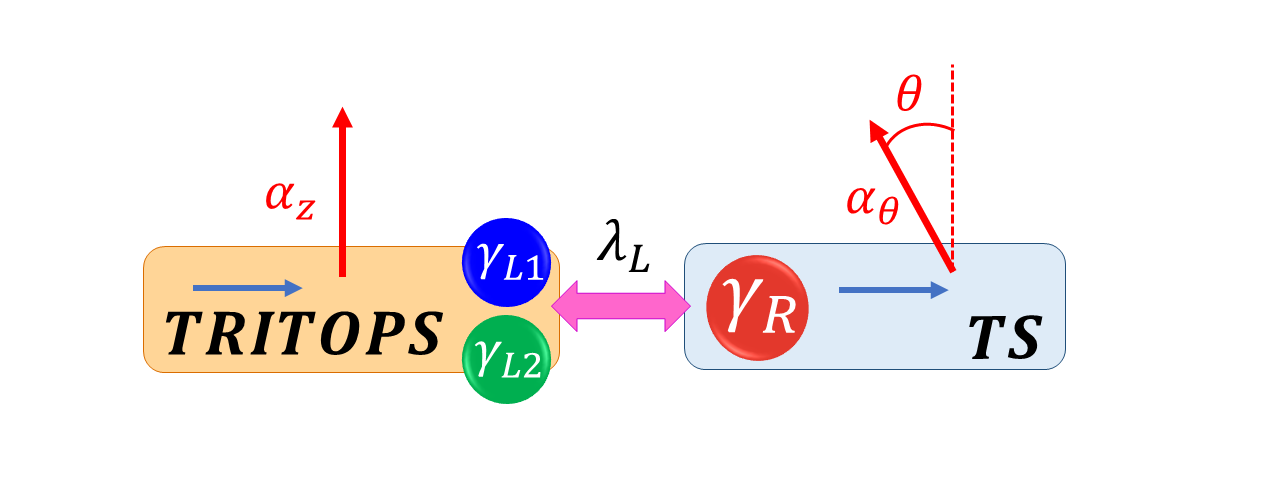}
\caption{Sketch of a TRITOPS-TS Josephson junction. Colored dots indicate MBSs corresponding to the Majorana operators $\gamma_{L1,L2,R}$. The tunnel coupling
$\lambda_{L}$ connects both wires, where blue arrows shows the positive momentum direction in each wire. The spin-orbit axes on both sides are tilted by the relative angle $\theta$.}
\label{fig10}
\end{figure}

We next consider the two-terminal Josephson junction in Fig.~\ref{fig10} between a TRITOPS wire [see Eq.~(\ref{kane_TRITOPS}) in Sec.~\ref{sec5a}], and a TS nanowire [see Eq.~\eqref{Lutchyn} in Sec.~\ref{sec3}].
Denoting the respective boundary spin-Nambu spinors by $\hat \Psi_L$ and $\hat \Psi_R$, respectively,
the tunneling Hamiltonian is given by
\be \label{htt}
H_T = \frac{1}{2} \lambda_{L} \hat{\Psi}^\dag_{L}\ \sigma_0 e^{i\tau_z\phi/2}\tau_z\hat{\Psi}_R^{} +{\rm  H.c.} ,
\ee 
where $\phi$ is the superconducting phase difference across the junction and we assume a real-valued tunnel coupling $\lambda_L$.  Below we assume for simplicity that the pairing gap $\Delta$ is identical for both nanowires. We will allow for a relative angle $\theta$ between the directions of the spin-orbit field in each wire, see the schematic device layout in
Fig.~\ref{fig10}. One could vary $\theta$ by changing the orientation of a local electric field applied to the TS wire only, which in turn will affect the corresponding Rashba spin-orbit field. 
In addition, we need a Zeeman field to induce the topological phase in the TS nanowire (see Sec.~\ref{sec3}), while no Zeeman field should be present on the time-reversal invariant TRITOPS side.
To achieve this goal, one may use mesoscopic ferromagnets for inducing a Zeeman field only locally \cite{Wei2019}.

\begin{figure}[t]
 \centering
  \includegraphics[width=\columnwidth]{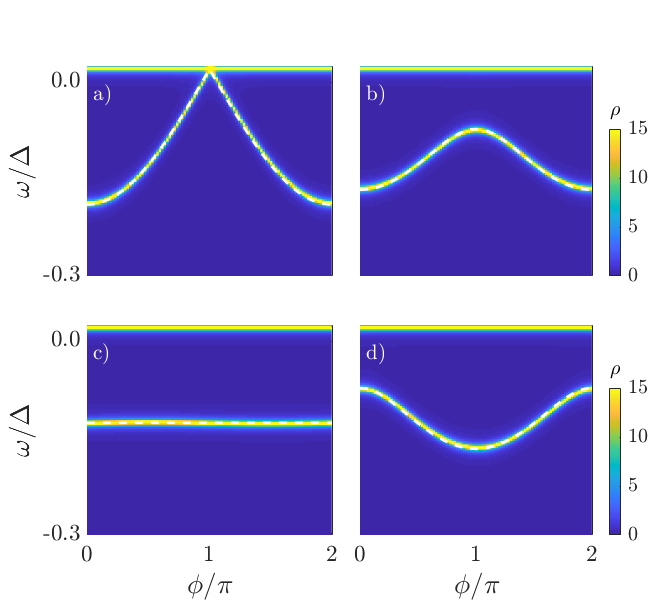}
\caption{Phase-dependent subgap spectrum of a TRITOPS-TS Josephson junction for different values of the tilt angle $\theta$
in Fig.~\ref{fig10}.
The spinful single-channel model parameters are as described in Secs.~\ref{sec3} and \ref{sec5}, with
$\mu = 1$~meV, $\lambda_L=2$~meV, and $V_x = 1.5V_c$ on the TS side.   The tilt angle is 
$\theta = 0$ in panel (a), $\theta= 0.3\pi$ in panel (b), $\theta= \pi/2$ in panel (c), 
and  $\theta=0.7\pi$ in panel (d). From blue to yellow, the color code indicates increasing DoS values at the junction, $\rho(\omega)$ (in meV$^{-1}$). 
White dashed curves show the Andreev bound states \eqref{abs3}.}
\label{fig11}
\end{figure}

To account for the angle $\theta$, we then apply the unitary transformation $R_y(\theta)$ to the bGF describing the TS nanowire. 
This rotation simultaneously affects the spin-orbit and the Zeeman field directions in the TS wire such that both directions can never be parallel
 to each other.
The junction spectral properties then follow again from a Dyson equation as in Eq.~(\ref{dyson-trijunction}).
Assuming that both wires have model parameters putting them deeply into the respective topological regime,
we can compare our numerically exact results for the subgap spectral properties 
to the corresponding predictions of a minimal model Hamiltonian.  The latter is obtained by retaining only the MBS degrees of freedom indicated in Fig.~\ref{fig10}.
To that end, the approximate expression for the boundary  spinors can again be derived from the respective bGFs as in Sec.~\ref{sec6a}. Those spinors involve the Majorana operators $\gamma_{L1,L2,R}$
in Fig.~\ref{fig10} and are given by 
\bea \nonumber
\hat{\Psi}_L &\simeq & \sqrt{\frac{\Delta}{t}} \bmat 1\\ 0 \\ i \\ 0 \emat \gamma_{L1} +  \sqrt{\frac{\Delta}{t}} \bmat 0\\ i \\ 0 \\ 1 \emat \gamma_{L2},\\
\hat{\Psi}_R &\simeq & \sqrt{\frac{\Delta_2}{t}} \, R_y(\theta) \bmat i\\ -i \\ 1\\1 \emat  \gamma_R.
\label{spinores_tritops_ts}
\eea 
The resulting minimal model Hamiltonian is 
\be 
H_{\rm min} = -i \left[w_1(\phi) \gamma_{L1}+w_2(\phi)\gamma_{L2}\right]\gamma_R
\label{minimal-model_tritops_ts}
\ee
with the energies 
\bea
w_{1}(\phi) &=&  \frac{2\lambda_L\sqrt{\Delta\Delta_2}}{t}\cos\frac{\phi}{2}\cos\frac{\theta}{2}, \nonumber \\
w_{2}(\phi) &=& -\frac{2\lambda_L\sqrt{\Delta\Delta_2}}{t}\sin\frac{\phi}{2}\sin\frac{\theta}{2} .  
\eea

The structure of $H_{\rm mm}$ in Eq.~\eqref{minimal-model_tritops_ts} is similar to the minimal model \eqref{tunneltriminimal} for the TS trijunction in Sec.~\ref{sec6a} without any 
coupling between the $\gamma_{L1,L2}$ operators.
The subgap spectrum is therefore characterized by a decoupled zero-energy Majorana state, and the hybridization of the two other Majorana operators yields  
the Andreev bound state dispersion:
\be 
E_\pm (\phi)= \pm \frac12 \sqrt{w_{1}^2(\phi)+ w^2_{2}(\phi) }.
\label{abs3}
\ee 
We compare Eq.~\eqref{abs3} to numerically exact results for the subgap spectral properties of the TRITOPS-TS junction in Fig.~\ref{fig11}.
Clearly, the general subgap spectrum is rather well described by the minimal model  (\ref{minimal-model_tritops_ts}). In contrast to the case of a tri-terminal TS junction, for TRITOPS-TS junctions
it is not necessary  to take into account higher-order tunneling processes for obtaining accurate  agreement with numerically exact  bGF calculations (but see 
Ref.~\cite{foot1}).

\section{Concluding remarks}
\label{sec7}

In the present work, we have generalized the  boundary Green's function approach of Refs. \cite{Zazunov2016,Zazunov2017} to quasi-1D spinful models of Majorana nanowires. 
For single-channel class D and class DIII wire models, we have obtained an analytical understanding of 
the behavior of the roots of the corresponding secular polynomial in complex momentum space.  This advance helps physical intuition and allows for a practical and numerically efficient method for computing the bGF, and thereby also physical observables.  
The method has also been extended to spinful multichannel models, where it appears to allow for more efficient numerical bGF calculations  than the alternative recursive technique \cite{Peng2017,Aligia2018}. Let us remark that the 
computational complexity of the method is only limited by 
the ability to evaluate the roots of a polynomial. Typically, the 
numerical demands are therefore much smaller than those for 
a recursive calculation of the bGF.

Given the efficient construction of the bGF put forward in this work, one can now apply the general bGF approach \cite{Zazunov2016} to study the transport properties of many different hybrid devices composed of Majorana nanowires 
and/or conventional metals or superconductor electrodes.  In Sec.~\ref{sec6}, we have provided two examples for such devices, namely phase-biased trijunctions of TS wires and TRITOPS-TS junctions.  
In both cases, we have carried out an analysis of the subgap Andreev (or Majorana) state dispersion at zero temperature.   

We believe that this approach offers many interesting perspectives for
future research.  In particular, one can study nonequilibrium transport properties away from the linear-response regime, and one can also include electron-electron or electron-phonon effects, at least on a
perturbative level.  We are confident that the results of our work can also be helpful for the interpretation of transport experiments carried out on hybrid devices containing nanowires with topologically nontrivial
superconducting phases.

\begin{acknowledgements}
We thank L. Arrachea and T. Martin for discussions. This project has been funded by the Deutsche Forschungsgemeinschaft under Germany's Excellence Strategy, Cluster of Excellence Matter
and Light for Quantum Computing (ML4Q) EXC 2004/1 - 390534769; by the Spanish Ministry of Economy, Industry, and Competitiveness through Grant No.~FIS2017-84860-R; and by the Mar{\'i}a de Maeztu Programme for Units of Excellence in n Research and Development Grant No.~MDM-2014-0377.
\end{acknowledgements}

\appendix
\section{On the spinful single-channel model}\label{appA}

In this appendix, we provide technical details pertaining to our discussion of the spinful single-channel Majorana wire
model \cite{Lutchyn2010,Oreg2010} in Sec.~\ref{sec2}.  First, 
the explicit form of the coefficients $C_n(\omega)$ in Eq.~(\ref{polynomial}) is given by
\begin{eqnarray}
C_0 &=& \frac{3 \alpha^4}{8} + \Delta^4 + \mu^4 - 8 \mu^3 t + 36 \mu^2 t^2 - 80 \mu t^3 \nonumber 
\\  & &  + 70 t^4 - 2 \mu^2 V_x^2 + 8 \mu t V_x^2 
 - 12 t^2 V_x^2  \nonumber \\
 & &+ V_x^4 - 2 (\mu^2 - 4 \mu t +  6 t^2  + V_x^2) \omega^2 + \omega^4 \nonumber \\
 & & + 2 \Delta^2 (\mu^2 - 4 \mu t  + 6 t^2 - V_x^2 - \omega^2)  \nonumber \\
 & & + \alpha^2 \left[ \Delta^2 - \mu^2 + 4 \mu t - 5 t^2 + V_x^2 - \omega^2\right] , \nonumber \\
 C_1 &=& -(\mu - 2 t) t (\alpha^2 - 
   4 (\Delta^2 + \mu^2 - 4 \mu t + 7 t^2  \nonumber \\ & &-V_x^2 - \omega^2)), \nonumber \\
   C_2 &=& \Bigl\{-\alpha^4 + 8 t^2 \Bigl(\Delta^2 + 3 \mu^2 - 12 \mu t + 14 t^2 - V_x^2 \nonumber \\
  & & - \omega^2\Bigr) + 2 \alpha^2 \left[-\Delta^2 + (\mu - 2 t)^2 - V_x^2 + \omega^2\right]\Bigr\}/4, \nonumber \\
  C_3 &=& (t\mu - 2t^2)(\alpha^2 + 4t^2), \quad
C_4 = \left[t^2 + \left (\alpha/2 \right)^2 \right ]^2 \quad 
\end{eqnarray}
It is convenient to renormalize these coefficients such that $C_4$ appears as common factor of the polynomial.

The $C_n$ coefficients in turn determine the coefficients $a_m(\omega)$ appearing in the eighth-order polynomial 
equation \eqref{8ord2}.  The roots $z_n(\omega)$ therefore have satisfy the Vieta relations 
\bea 
S_k(z_1,\ldots,z_8) &=& \sum_{i_1<i_2<\cdots<i_k} z_{i_1}z_{i_2}\cdots z_{i_k} \nonumber \\ &=& (-1)^{8-k}\frac{a_k}{a_{8}}.
\eea 
Using the condition \eqref{conditionadd} and the ansatz \eqref{ansatz}, the first three invariants are given by
$S_1=2AB$, $S_2=2(A^2-2)(1+C)+4B^2$, and $S_3 = 2AB(A^2-1)+4ABC$ with
\begin{eqnarray}
A &=& 1-\eta + \frac{1}{1-\eta},  \quad
B = \cos(\delta_1) + \cos(\delta_2), \nonumber \\
C &=& 2\cos(\delta_1)\cos(\delta_2). 
\label{ABC}
\end{eqnarray}
As a consequence, the parameter $C$ obeys a cubic equation that can be solved analytically,
\be\label{cubiceq}
w_3 + w_2C + w_1C^2 + C^3 = 0,
\ee
with the coefficients
\bea 
w_1 &=& 1-\frac{S_3}{S_1},\nonumber  \\
w_2 &=& \frac{S_2}{4} -\frac{S_3}{S_1} -\frac{1}{4} + \left(\frac{S_3}{2S_1}\right)^2, \nonumber \\
w_3 &=& -\frac{S_2S_3}{8S_1} - \frac{S_2}{8} + S_1^2-\frac{1}{4} + \left(\frac{S_3}{2S_1}\right)^2.
\eea 
For $V_x<V_c$, the physical solution of Eq.~\eqref{cubiceq} is given by
\be
C = -2 \sqrt{-Q}\cos(\theta_0/3) -w_1/3,
\ee
with 
\bea 
\theta_0 &=& \cos^{-1}\left(-\frac{R}{\sqrt{-Q^3}}\right), \nonumber  \\
Q &=& \frac{3a_2-w_1^2}{9}, \nonumber \\ 
R &=& \frac{9w_1w_2 - 27w_3 - 2w_1^3}{54}. 
\eea 
For $V_x>V_c$, the solution is given by 
$C = P_1-Q/P_1 - w_1/3$ (assuming $P_1\neq 0$), with
\be 
P_1 = \text{sgn}(R)\left( |R| + \sqrt{R^2+Q^3} \right)^{1/3}.
\ee
The coefficients $A$ and $B$ then follow from
\be 
A^2=\frac{S_3}{S_1} + 1 -2C, \quad B=\frac{S_1}{2A}. 
\ee
Finally, the parameters in our ansatz [see Eqs.~\eqref{ansatz} and \eqref{ansatz2}], can be determined from the relations
\bea 
\cos\delta_1 &=& \frac{B+\sqrt{B^2-2C}}{2}, \nonumber \\
\cos\delta_2 &=& \frac{B-\sqrt{B^2-2C}}{2}, \nonumber \\  
\eta &=& 1-\frac{A}{2}+\sqrt{\frac{A^2}{4}-1} . 
\label{parameters}
\eea

We proceed  by providing the detailed form of the matrices $\hat{A}_{\nu}$ and $\hat{A}'_{\nu}$ in Eq.~(\ref{approximate-GF}).
Using the definition in the main text, for $V_x<V_c$, they are with $z_\nu(\omega)$ in Eq.~\eqref{ansatz} given by
\be \label{bbb1}
\hat{A}_{\nu} = \frac{\hat{A}(z_\nu)}{b_\nu} + \frac{\hat{A}(z^*_\nu)}{b^*_\nu}, \quad
\hat{A}^\prime_{\nu} = \frac{\hat{A}'(z_\nu)}{b_\nu} + \frac{\hat{A}'(z^*_\nu)}{b^*_\nu},
\ee
where an expansion of $\prod_{z_\nu\neq z_m}(z_\nu-z_m)$ to first order in $\tau_\nu\sqrt{\Delta^2_\nu - \omega^2}$ yields
\be 
b_\nu = 32e^{3i\delta_\nu}\tau_\nu\sin^2(\delta_\nu)\left[\cos({\delta_2})-\cos({\delta_1}) \right]^2.
\ee 
Explicitly, the components of the symmetric $4\times 4$ matrix $\hat A$ in Eq.~\eqref{bbb1}, $\hat{A}_{ij}=\hat{A}_{ji}$, follow from
\bea 
\hat{A}_{11}(z) &=& -\hat{A}_{33}(z) =z^3 V_x^2\left[\epsilon(z)-\tilde\alpha(z)\right]  \nonumber\\ \nonumber
&+&z^3[\epsilon(z)+\tilde\alpha(z)]\left\{-\Delta^2 - (\epsilon(z)- \tilde\alpha(z))^2\right\}, \nonumber \\ \nonumber
\hat{A}_{22}(z) &=& - \hat{A}_{44}(z)=z^3V_x^2\left[\epsilon(z)+\tilde\alpha(z)\right] \\
&+&
z^3[\epsilon(z)-\tilde\alpha(z)]\left\{-\Delta^2 - (\epsilon(z)+\tilde\alpha(z))^2\right\}, \nonumber \\ 
\hat{A}_{12}(z) &=& \hat{A}_{34}(z)= z^3V_x\left[\Delta^2 + \epsilon^2(z) - \tilde\alpha^2(z) -V_x^2  \right],\nonumber \\
\hat{A}_{13}(z) &=& z^3 \Delta \left\{V_x^2 - \Delta^2- [\epsilon(z)-\tilde\alpha(z)]^2\right\}, \nonumber \\
\hat{A}_{14}(z) &=& -\hat{A}_{23}(z) = 2z^3V_x\Delta \tilde\alpha(z),\nonumber\\
\hat{A}_{24}(z) &=& z^3 \Delta \left\{V_x^2 - \Delta^2- [\epsilon(z)+\tilde\alpha(z)]^2\right\}. 
\eea 
Similarly, by taking a derivative with respect to $\omega$,  the nonvanishing matrix elements of the symmetric matrix $\hat{A}^\prime_{ij}=\hat{A}^\prime_{ji}$  follow as
\bea 
\hat{A}^\prime_{11}(z) &=& \hat{A}^\prime_{33}(z)=-z^3\left\{\Delta^2 +[\epsilon(z)-\alpha(z)]^2+V_x^2\right\},\nonumber \\
\hat{A}^\prime_{22}(z) &=& \hat{A}^\prime_{44}(z)=-z^3\left\{\Delta^2 + [\epsilon(z)+\alpha(z)]^2+V_x^2\right\}, \nonumber \\
\hat{A}^\prime_{12}(z) &=& -\hat{A}^\prime_{34}(z)= 2z^3 V_x \epsilon(z), \nonumber \\
\hat{A}^\prime_{14}(z) &=& \hat {A}^\prime_{23}(z)= 2z^3 V_x\Delta.
\eea
In the topologically nontrivial phase, $V_x>V_c$,  trigonometric functions 
associated with the roots $z_{1,\pm}$ in Eq.~\eqref{ansatz2} turn into hyperbolic functions. 
The matrices with $\nu=1$ in Eq.~\eqref{bbb1} are then replaced by   
\bea 
\hat{A}_{1} &=& \frac{\hat{A}(z_{1,+})}{\tilde b_1} - \frac{\hat{A}(z_{1,-})}{\tilde b_1}, \nonumber \\
\hat{A}_{1}^\prime &=& \frac{\hat{A}^\prime(z_{1,+})}{\tilde b_1} - \frac{\hat{A}^\prime(z_{1,-})}{\tilde b_1},
\eea
with the quantities  
\bea 
\tilde b_1 &=& 32e^{-3\delta_1}\tau_1\sinh^2(\delta_1)\left[\cos(\delta_2)-\cosh(\delta_1)\right]^2, \nonumber \\
\tilde b_2 &=& 32e^{3i\delta_2}\tau_2\sin^2(\delta_2)\left[\cos(\delta_2)-\cosh(\delta_1) \right]^2. 
\label{residues}
\eea 
The $\nu=2$ matrices follow from  Eq.~\eqref{bbb1} with the replacement $b_2\to \tilde b_2$.
Finally, we note that for very large $V_x$,  one approaches the Kitaev limit of the nanowire, and the relevant residues come from the $z_2$ roots only.

\section{On TRITOPS wires}\label{appB}

According to Theorem 1 of Ref.~\cite{Mong2011}, the largest-modulus root $z_{\rm max}$ inside the unit circle
 can be determined from the relative position of the origin inside the ellipse discussed in Sec.~\ref{sec5a}.
For that purpose, we first determine the major ($M$) and minor $(m)$ axes of the ellipse in Fig.~\ref{fig5}.
Using Eq.~(\ref{beta}) and focusing on the case of $\hat{\cal H}_-(k)$, the defining equation of the ellipse is given by
\be \label{B1}
B^T\left(\begin{array}{cc} (t^2+\alpha^2)/\Delta^2 & t/\Delta \\ t/\Delta & 1 \end{array}\right)
B = 4\alpha^2
\ee 
with $B^T=(\beta_{-,x}, \beta_{-,z}+\mu)$.
From the eigenvalues of the $2\times 2$ matrix in Eq.~\eqref{B1},
\be 
\lambda_\pm = \frac{t^2+\Delta^2+\alpha^2}{2\Delta^2}\pm \frac{\sqrt{(t^2+\Delta^2)^2+2(t^2-\Delta^2)\alpha^2+\alpha^4}}{2\Delta^2},
\ee 
we obtain 
\be\label{mM}
m = 4\alpha/\sqrt{\lambda_+}, \quad M = 4\alpha/\sqrt{\lambda_-}.
\ee 
The distance between the foci of the ellipse then follows as $f = \sqrt{M^2-m^2}$. 

To obtain the distance $l = |OF_1|+|OF_2|$ between the foci and the origin (corresponding to the red dashed line in Fig.~\ref{fig5}),  
we first compute the rotation angle $\theta$ of the ellipse using the eigenvectors of the conic section matrix,
\bea 
\cos\theta &=& \frac{1}{\sqrt{1 + X^2/(2t\Delta)^2}} ,\nonumber \\
 X&=& \Delta^2-t^2-\alpha^2 \nonumber \\  & -&\sqrt{(t^2+\Delta^2)^2+2(t^2-\Delta^2)\alpha^2+\alpha^4} .
\eea
As a consequence, $l$ follows from the relation
\be 
|OF_{1,2}| = \sqrt {(f/2)^2 + \mu^2 \pm \mu f \sin\theta} .
\ee
The largest-modulus root inside the unit circle is then given by (see Refs.~\cite{Mong2011,Aligia2018})
\be \label{zmax-tritop}
|z_{\rm max}| = \frac{l + \sqrt{l^2 - f^2}}{M + m}.
\ee 
The same result follows for the other block, $\hat{\cal H}_+(k)$.
As discussed in Sec.~\ref{sec5a}, Eq.~\eqref{zmax-tritop} determines the decay length of Majorana end states into the bulk of a TRITOPS wire.

\end{document}